\documentclass{aa520}
\usepackage{natbib}
\usepackage{times,graphics,latexsym,amssymb}
\bibpunct{(}{)}{;}{a}{}{,}
\begin{document}

\title{Stochastic chemical enrichment in metal-poor systems}
\subtitle{I. Theory}
\author{T. Karlsson}
\offprints{T. Karlsson (Torgny.Karlsson@astro.uu.se)}
\institute{Department of Astronomy and Space Physics, Box 515, SE-751 20, Uppsala, Sweden}
\titlerunning{Stochastic chemical enrichment}
\abstract{
\noindent
A stochastic model of the chemical enrichment of metal-poor systems by core-collapse (Type~II) supernovae is presented, allowing for large-scale mixing of the enriched material by turbulent motions and cloud collisions in the interstellar medium. Infall of pristine material is taken into account by following the evolution of the gas density in the medium. Analytical expressions were derived for the number of stars enriched by a given number of supernovae, as well as for the amount of mass with which the ejected material from a supernova is mixed before being locked up in a subsequently formed star. It is shown that for reasonable values of the gas density ($\sim0.1$ cm$^{-3}$) and of the supernova rate ($\sim 0.25$ kpc$^{-3}$ Myr$^{-1}$) of the Galactic halo, the resulting metallicity distributions of the extreme Population II stars show a distinct cut-off at $[\mathrm{Fe}/\mathrm{H}]\simeq -4$. In fact, by assuming no low-mass Population III stars were able to form out of the primordial interstellar medium, the derived fraction of stars below $[\mathrm{Fe}/\mathrm{H}]= -4$ is in agreement with observations. Moreover, the probability is high that even the most metal-poor stars observed to date have been enriched by several contributing supernovae. This partly explains the relatively small star-to-star scatter in many chemical-abundance ratios for stars down to $[\mathrm{Fe}/\mathrm{H}]= -4$, as recently found in several observational studies. Contribution from the thermonuclear (Type~Ia) supernovae is found to be negligible over almost the entire extremely metal-poor regime. Although the fraction of contaminated stars may increase rapidly towards $[\mathrm{Fe}/\mathrm{H}]= -2.5$, the fraction of stars with iron primarily from Type~Ia supernovae remains small. The stars that are heavily polluted by Type~Ia supernovae are pushed towards higher metallicities, creating a hole/bump in the metallicity distribution. Such features could be used to reveal the possible presence of subpopulations of Galactic halo stars that have been enriched by Type~Ia supernovae.    
}
\maketitle

\section{Introduction} \label{intro}
\noindent
It is an interesting fact that in the extremely metal-poor regime ($[\mathrm{Fe}/\mathrm{H}]<-2.5$) of the Galactic halo, the observed star-to-star scatter in abundance ratios of elements heavier than helium tend to be smaller than previously anticipated (Carretta et al.\ 2002\nocite{carretta02}; Cayrel et al.\ 2004\nocite{cayrel04}), while in the metal-rich regime ($[\mathrm{Fe}/\mathrm{H}]>-1$) of the Galactic disk, the scatter seems larger than one would expect from simple stochastic enrichment considerations (e.g., Edmunds 1975\nocite{edmunds75}). As noted by, e.g., Cayrel et al.\ (2004\nocite{cayrel04}), the dispersion is surprisingly small in many abundance ratios, excluding the neutron-capture elements, even for stars with a metallicity as low as $[\mathrm{Fe}/\mathrm{H}]=-4$. Supposedly (see, e.g., Woosley \& Weaver 1995\nocite{ww95}), the stellar yields of different heavy nuclei depend strongly, among other things, on the progenitor mass of the supernova (SN). It therefore seems that a small scatter can only be obtained if these stars were formed out of gas enriched by several SNe (cf. Karlsson \& Gustafsson 2005, hereafter Paper~II\nocite{kg04}). 

\par

Could it be that averaging effects were already in play at a metallicity of $[\mathrm{Fe}/\mathrm{H}]\sim-4$? This question leads us to another: What are the number and the metallicities of stars enriched by any given number of SNe? Several chemical evolution models have been developed to study the effect of small number statistics on the build-up of chemical elements in the early Galaxy (e.g., Tsujimoto et al.\ 1999\nocite{tsy99}; Argast et al.\ 2000\nocite{argast00}; Oey 2000\nocite{oey00}; Travaglio et al. 2001\nocite{travaglio01}). However, these models do not explicitly trace or discuss the metallicity distributions of individual populations of stars enriched by a specific number of SNe. In the simplest case one might assume that the metallicity is proportional to the number of enriching SNe (e.g., Fields et al.\ 2002\nocite{fields02}); but as we shall see, allowing a variable dilution (increasing with time) of the SN material and a variable stellar yield introduces a dispersion in the number of SNe contributing heavy elements to stars, even for stars with identical metallicity. Moreover, due to the continuous mixing of the interstellar medium (ISM) and the possible infall of primordial gas, stars of a given metallicity will, on average, be enriched by more SNe than expected from such simple considerations.      

\par

The interesting approach taken by Oey (2000\nocite{oey00}) examined stochastic enrichment of the ISM by randomly distributed star-forming regions, where each region was assigned a specific metallicity depending on its predetermined size. A similar approach was used here. However, we focus instead on the SN remnants themselves, which allows us to study the effects of individual stars on the chemical enrichment of the ISM. A more detailed comparison between the present approach and that of Oey (2000\nocite{oey00}) will be given in Sect.~\ref{results_procon}.  

\par

In Sect.~\ref{theory} the stochastic theory is outlined, including a derivation of the expression for the number of stars enriched by a certain number of SNe, as well as an expression for the amount of dilution of the newly synthesized material. A stochastic enrichment model for metal-poor systems is then presented in Sect.~\ref{stoch}, making use of these results. In Sect.~\ref{application} three specific models are presented, applicable to the Galactic halo, and in Sect.~\ref{results} the results are discussed. A summary is given in Sect.~\ref{conclusions}.

\section{The chemical evolution of the ISM}\label{theory}
\noindent
The chemical enrichment of the ISM proceeds as the material ejected in SN explosions, planetary nebulae, and stellar winds disperses and gradually mixes with the ambient medium. Due to the non-instantaneous mixing, local enrichment events, such as SNe, cause the interstellar abundance $A/\mathrm{H}$ of an element $A$ relative to hydrogen to vary over space and time. This variation can be described by a function $f_A(t,\mathbf{x})$, which is different for each element. In order to link the evolution of chemical inhomogeneities in the ISM with the abundances observed in low-mass stars, we require the following knowledge: what and how much of an element is ejected from each SN; when, where, and at what rate SNe explode; how the material is dispersed; and finally, when and where the local abundance is probed by subsequent star formation. Mapping of $f_A(t,\mathbf{x})$ onto the population of low-mass stars is controlled and weighted by the star formation rate (SFR) $\psi=\psi(t,\mathbf{x})$, which is thought to favour cool, dense, and gravitationally unstable regions of the ISM. This last step contains an integration over space and time and results in abundance distributions describing the total probability of finding a star with a certain set of chemical abundances. Such distributions can be compared with the observations.

\par

Modelling the evolution of chemical elements with intrinsic fluctuations mapped by the low-mass stars is an intricate task. Even extensive numerical simulations suffer from simplified physics and inadequate spatial resolution. In this study, we instead discuss an analytical approach that neglects the detailed physics of star formation and treats the discrete chemical enrichment as a stochastic process.

\subsection{Properties of the model ISM}\label{theory_modelism}
\noindent
Let us first write down some definitions. The probability that a newly formed star has a mass $m$ is given by the initial mass function (IMF), which is assumed to be a time-independent random variable with probability density function \mbox{$f_m\equiv \phi(m)$}, normalized by number such that

\begin{equation}
\int\limits_{m_l}^{m_u}\phi(m)\mathrm{d}m = 1.
\label{imf}
\end{equation}

\noindent
The lower and upper mass limits are denoted by $m_l$ and $m_u$, respectively. 

\par

In this study, we shall mainly discuss the early chemical evolution of primary elements---like oxygen, magnesium, and iron---produced in core-collapse SNe (i.e., mainly Type~II SNe; however, see Sect. \ref{sectsnia} for a discussion of the effect of thermonuclear SNe). The total mass of such an element $A$, produced in the progenitor star and in the explosion, is predominantly determined by the mass of the star and shall be denoted by $p_{A}(m)$. The mass fraction of $A$ in the gas from which the star was formed should be taken into account formally when calculating the total amount of mass $e_{A}$ ejected in the explosion. However, at low metallicities, say for $[\mathrm{Fe}/\mathrm{H}]\le -2$, the ejected amount of $A$ is completely dominated by the yield $p_{A}$. Thus, we will assume that $e_{A}=p_{A}(m)$. 

\par

Star formation is a clustered phenomenon, and many stars are formed in groups like globular clusters, open clusters, and OB associations. However, there is also less clustered star formation, and most of the field stars cannot have been formed in strong gravitationally bound associations. A fundamental assumption in the present model is that star formation is unclustered and that the stars are randomly distributed in space. This means that the SFR can be described by a function $\psi=\psi(t)$ that depends only on time. Under this assumption, chemical enrichment events, such as SNe, are spatially uncorrelated, as is mapping of the chemical inhomogeneities. Within the present theory this assumption may, however, be replaced by the assumption of correlated star formation in stochastically independent star-forming regions (Karlsson 2005, hereafter Paper~III\nocite{paperiii}). 

\par
  
Like the SFR, the density $\rho(t,\mathbf{x})$ in the model ISM is described by the spatial average $\rho(t)$. The density changes over time due to three possible mechanisms: infall of gas onto the system, outflow of gas, and star formation. The rate of change is governed by

\begin{equation}
\frac{\mathrm{d}\rho}{\mathrm{d}t} = u_{\mathrm{acc}}-u_{\mathrm{ej}}-\frac{\mathrm{d}s}{\mathrm{d}t},
\label{drhodt}
\end{equation}    

\noindent
where $u_{\mathrm{acc}}$ is the accretion rate, $u_{\mathrm{ej}}$ the ejection rate, and $\mathrm{d}s/\mathrm{d}t$ the rate of mass locked up in stars and stellar remnants, which is given by the amount of mass in newly formed stars minus the amount of mass returned to the ISM per unit time by the dying stars. The lifetime $\tau$ of a star depends mainly on its mass and will therefore be denoted by the function $\tau=g_{\tau}(m)$. Using this notation, the stars of mass $m$ returning their outer layers to the ISM at time $t$ were formed at a rate $\psi(t-g_{\tau}(m))$ per unit volume. Moreover, the mass of the least-massive dying star depends on time such that $m_l(t)=g_{\tau}^{-1}(t)$. Hence,

\begin{equation}
\frac{\mathrm{d}s}{\mathrm{d}t} = \langle m \rangle \psi(t)\,\,
- \!\!\int\limits_{g_{\tau}^{-1}(t)}^{m_u}\!\!(m-m_r)\phi(m)\psi(t-g_{\tau}(m))\mathrm{d}m,
\label{dsdt}
\end{equation}  

\noindent
where $\langle m \rangle=\int_{m_l}^{m_u}m \phi(m)\mathrm{d}m$ is the average mass of stars in a stellar generation. Note that mass $m_r$ of the stellar remnant is also a function of $m$. Given the infall- and outflow rates, together with an initial value $\rho_0$, $\rho(t)$ is fixed by Eq.~(\ref{drhodt}). The gas density will be used in calculating the amount of dilution of the newly synthesized elements (ejected in the SN explosions) before they are locked up in subsequently formed stars (Sect.~\ref{theory_mixingmass}).

\par

The way infall is introduced in Eq.~(\ref{drhodt}) implies that the mixing of infalling gas with the ISM is instantaneous. The infall is assumed to be a homogeneous, diffuse flow of gas onto the system and not in the form of, e.g., discrete clouds. We will also suppose that the infalling gas is pristine, consisting of pure hydrogen and helium, which implies that any chemical inhomogeneities produced by our model originate and evolve within the system. Thus, evolution beyond the system is not considered. 

\par

\begin{figure}[t]
 \resizebox{\hsize}{!}{\includegraphics{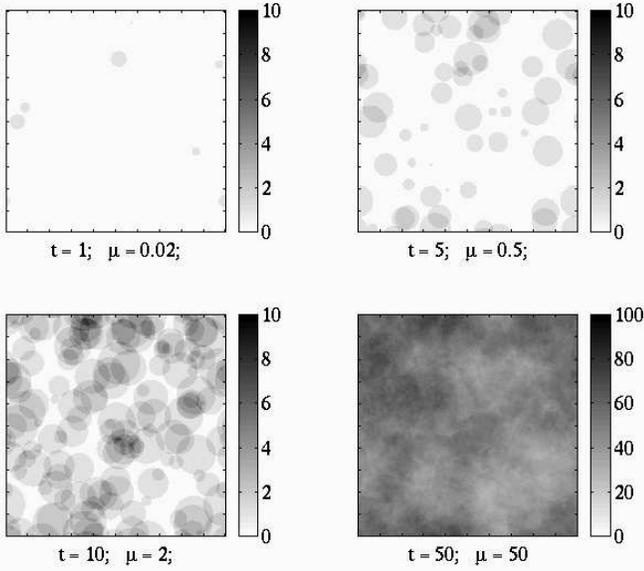}}
 \caption{Illustration of the gradual increase in the volume containing material from SNe. The time-like parameter $\mu$, defined in Eq.~(\ref{muoft}), is a measure of the average number of SNe that have enriched a volume element. Here, $\mu \propto t^2$. The SNe are randomly distributed over space at a constant rate $u_{\mathrm{SN}}$. The shaded bar indicates the number of enriching SNe. Time is expressed in arbitrary units.}
 \label{sfr_stat}
\end{figure}

\subsection{The mixing volume picture} \label{mixing_volume}
\noindent
The total volume enriched by debris from a single SN shall be defined as the mixing volume $V_{\mathrm{mix}}$ of that SN. The mixing volume will grow steadily with time, initially due to expansion of the SN remnant and later on due to large-scale mixing processes, such as turbulent diffusion. In this picture a mixing volume is generated by each SN event. Chemical inhomogeneities develop with subsequent generations of SNe that explode within earlier generations of mixing volumes, so that mixing volumes begin to overlap and the gas becomes further enriched. This scenario is illustrated in Fig.~\ref{sfr_stat}. At any instant in time, different regions of the ISM will be enriched by different numbers of SNe, and a distribution $w_{\mathrm{ISM}}(k,t)$ exists to describe the probability of finding $k$ overlapping mixing volumes at time $t$ somewhere in the ISM. Knowing this distribution, we are able to quantify and map the abundance distributions in stars.

\par

\begin{figure}[t]
 \resizebox{\hsize}{!}{\includegraphics{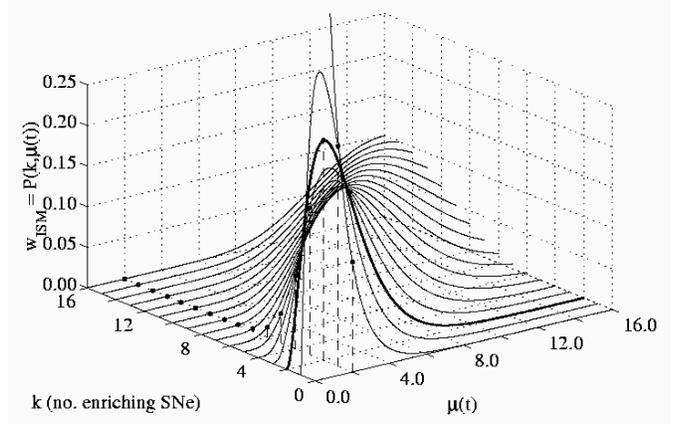}}
 \caption{The two-dimensional probability density function $w_{\mathrm{ISM}}(k,t)$ as given by Eq.~(\ref{poisson}). In particular, the thick line gives the probability of finding a region in space enriched by $2$ SNe as a function of the time-like variable $\mu$. The dots give the probability of finding a region enriched by $k$ SNe at $\mu=2.0$ (cf. Fig.~\ref{sfr_stat}). Note that $w_{\mathrm{ISM}}(0,t)$ is the probability of finding an unenriched region.}
 \label{wism_fig}
\end{figure}

\subsection{The distribution of chemical inhomogeneities} \label{theory_distr}
\noindent
To find a general expression for $w_{\mathrm{ISM}}(k,t)$ for a turbulent and inhomogeneous ISM with a spatially clustered SN distribution is beyond the scope of this study. However, if the SNe are randomly distributed over space with an average rate $u_{\mathrm{SN}}$ per unit time and unit volume, we take the probability of finding a region enriched by $k$ SNe at time $t$ to be approximated by the Poisson distribution (see Fig.~\ref{wism_fig}), i.e.,

\begin{equation}
w_{\mathrm{ISM}}(k,t)=P(k,\mu(t))=e^{-\mu(t)}\mu(t)^{k}/k!,
\label{poisson}
\end{equation}

\noindent
where $\mu(t)$ is the integrated volume affected by SNe, expressed in units of the total volume of the system at time $t$. Dispersive processes, like turbulent diffusion, will gradually spread the SN material over larger and larger volumes. A mixing volume that grows with time would mimic such an effect. More precisely, if $V_{\mathrm{mix}}(t-t')$ is the size of the volumes at time $t$ enriched by material from SNe that exploded at time $t'$, then 

\begin{equation}
\mu(t) = \int\limits_{0}^{t} V_{\mathrm{mix}}(t-t')u_{\mathrm{SN}}(t')\mathrm{d}t'.
\label{muoft}
\end{equation}

\noindent
With this definition, $\mu(t)$ is a measure of the average number of SNe that have enriched a random volume element in space at time $t$. By introducing a function of the form $\theta(t-t')\theta(t')$, where $\theta$ is the Heaviside step function, we can formally extend the limits of integration to $\pm \infty$, thus expressing $\mu$ as a convolution such that

\begin{eqnarray}
\mu(t) & = & \int\limits_{-\infty}^{\infty} V_{\mathrm{mix}}(t-t')\theta(t-t')\theta(t')u_{\mathrm{SN}}(t')\mathrm{d}t' \nonumber\\
& \equiv & \mathcal{V_{\mathrm{mix}}}\ast \mathcal{\scriptstyle U}_{\mathrm{SN}}(t),
\label{muconv}
\end{eqnarray}

\noindent
where $\mathcal{V}_{\mathrm{mix}}=V_{\mathrm{mix}}(t)\theta(t)$ and $\mathcal{\scriptstyle U}_{\mathrm{SN}}=u_{\mathrm{SN}}(t)\theta(t)$. This result will be used in Sect.~\ref{theory_counts} to derive an expression for the total number density of stars enriched by a certain number $k$ of SNe.

\subsection{Star counts} \label{theory_counts}
\noindent
Even though the increase in the average abundance of an element in the ISM can be considered as a continuous evolution, the local abundance increases due to discrete enrichment events, such as SNe. Therefore, it might be more natural to designate the build-up of heavy elements in the ISM as a chemical enrichment rather than a chemical evolution, especially in the extremely metal-poor regime. We emphasize this distinction in our treatment by integrating over time and deriving expressions with the number $k$ of enriching SNe as the explicit variable.

\par 

Let us first calculate the total number of stars enriched by a given number $k$ of SNe assuming that the probability of finding a region enriched by $k$ SNe at time $t$ is given by Eq.~(\ref{poisson}). If $\psi(t)$ is the SFR per unit time and unit volume, the number density of stars formed in regions enriched by $k$ SNe in the time interval $[t,t+\mathrm{d}t]$ is $w_{\mathrm{ISM}}(k,t)\psi(t)\mathrm{d}t$. The total number of still-existing stars per unit volume formed after a long time is then given by the integral

\begin{equation}
n_{k} = \int\limits_{0}^{\tau_\mathrm{G}} a_{\star}(t)w_{\mathrm{ISM}}(k,t) \psi(t)\mathrm{d}t,
\label{Nk}
\end{equation}

\noindent
where $\tau_{\mathrm{G}}=13.5$ Gyr is the age of the oldest stars (Bennett et al.\ 2003\nocite{bennett03}), which we assume coincides with the age of the Galaxy, and $a_{\star}(t)$ is the fraction of a stellar generation formed at $t$ that still exists today. The latter is given by

\begin{equation}
a_{\star}(t)=\int\limits_{m_{l}}^{g_{\tau}^{-1}(\tau_{\mathrm{G}}-t)}\!\!\!\!\!\!\!\!\!\phi(m)\mathrm{d}m, 
\label{aexist}
\end{equation} 

\noindent
where $\phi(m)$ is the IMF and $g_{\tau}^{-1}(\tau_{\mathrm{G}}-t)$ denotes the mass of the most massive, still-surviving stars formed at time $t$. Note, however, that in the early Galaxy, when most of the stars with [Fe$/$H] $<-2.5$ were formed, $a_{\star}$ changes very little. Therefore, in the extremely metal-poor regime we may approximate the integral in Eq.~(\ref{aexist}) with a constant such that $a_{\star}=a_{\mathrm{LMS}} $, where $a_{\mathrm{LMS}}$ is the fraction of low-mass stars with lifetimes greater than the age of the Galaxy.

\par

A closed expression for $n_k$ cannot be obtained in the general case. However, by the change of variables $t\rightarrow \mu$ we can approximate the integral in Eq.~(\ref{Nk}) by a function $\mathcal{\scriptstyle N}$ and its derivatives. To find $\mathcal{\scriptstyle N}$, we first need an expression for $\mathrm{d}\mu/\mathrm{d}t$, which is given by the derivative of the expression in Eq.~(\ref{muconv}). The derivative of a convolution is formally given by $\frac{\mathrm{d}}{\mathrm{d}t}(\mathcal{V_{\mathrm{mix}}}\ast \mathcal{\scriptstyle U}_{\mathrm{SN}})=\dot{\mathcal{V}}_{\mathrm{mix}}\ast \mathcal{\scriptstyle U}_{\mathrm{SN}}~(\equiv \mathcal{V}_{\mathrm{mix}}\ast \dot{\mathcal{\scriptstyle U}}_{\mathrm{SN}})$ where the derivative of $\mathcal{V_{\mathrm{mix}}}$ is 

\begin{equation}
\dot{\mathcal{V}}_{\mathrm{mix}}=\frac{\mathrm{d}}{\mathrm{d}t}(V_{\mathrm{mix}}\theta)=\dot{V}_{\mathrm{mix}}\theta+V_{\mathrm{mix}}\delta,
\label{vdot}
\end{equation} 

\noindent
and $\delta$ is the Dirac delta function. Since $V_{\mathrm{mix}}(0)=0$, the second term in Eq.~(\ref{vdot}) vanishes in the integration, and we obtain 

\begin{equation}
\frac{\mathrm{d}\mu}{\mathrm{d}t} = \dot{\mathcal{V}}_{\mathrm{mix}}\ast \mathcal{\scriptstyle U}_{\mathrm{SN}}
= \int\limits_{0}^{t} \dot{V}_{\mathrm{mix}}(t-t')u_{\mathrm{SN}}(t')\mathrm{d}t'.
\label{dmudt1}
\end{equation}

\noindent
The SN rate $u_{\mathrm{SN}}(t)$ relates to $\psi(t-\tau_{\mathrm{SN}})$, where $\tau_{\mathrm{SN}}$ is the typical lifetime of a high-mass star that explodes as an SN. However, neglecting this small time delay between the SFR and the SN rate and assuming a time-independent IMF, $u_{\mathrm{SN}}$ is proportional to $\psi$ such that 

\begin{equation}
u_{\mathrm{SN}}(t)=a_{\mathrm{SN}}\psi(t),
\label{asnpsi}
\end{equation}

\noindent
where $a_{\mathrm{SN}}$ is the fraction of high-mass stars in a stellar generation. This assumption is used in all the calculations that follow, and for Eq.~(\ref{dmudt1}) we get 

\begin{equation}
\frac{\mathrm{d}\mu}{\mathrm{d}t} = a_{\mathrm{SN}}\int\limits_{0}^{t} \dot{V}_{\mathrm{mix}}(t-t')\psi(t')\mathrm{d}t' \equiv a_{\mathrm{SN}}\mathcal{R}, 
\label{dmudt2}
\end{equation}

\noindent
where the integral denoted by $\mathcal{R}=\mathcal{R}(\mu(t))$ is a measure of the rate of change of $\mu$.

\par  

Since both $\dot{V}_{\mathrm{mix}}$ and $u_{\mathrm{SN}}$ are, by necessity, non-negative functions of time, $\mu(t)$ increases monotonically and $t$ can therefore be expressed in terms of $\mu$. Thus, the change of variables $t \rightarrow \mu$ yields $\mathrm{d}\mu/\mathrm{d}t=a_{\mathrm{SN}}\mathcal{R}(\mu)$ and $\psi=\psi(\mu)$, so that Eq.~(\ref{Nk}) becomes

\begin{equation}
n_{k} = \frac{1}{k!}\int\limits_{0}^{\mu_{\mathrm{G}}} e^{-\mu}\mu^k \frac{\psi/\mathcal{R}}{a_{\mathrm{SN}}/a_{\mathrm{LMS}}}\mathrm{d}\mu,
\label{NkR1}
\end{equation}

\noindent
where $\mu_{\mathrm{G}}=\mu(\tau_{\mathrm{G}})$. Recall that for the early Galaxy, \mbox{$a_{\star}\simeq a_{\mathrm{LMS}}$}. As soon as $w_{\mathrm{ISM}}$ starts to decrease, it quickly goes to zero due to the exponential $e^{-\mu}$ (see Fig.~\ref{wism_fig}). This means that for relatively small $k$ (i.e., at low metallicities, depending on the value of $\mu_{\mathrm{G}}$) the upper integration limit $\mu_{\mathrm{G}}$ can, in practice, be extended to $+\infty$. Identifying the function $\mathcal{\scriptstyle N}$ with

\begin{equation}
\mathcal{\scriptstyle N}(\mu) \equiv \frac{\psi/\mathcal{R}}{a_{\mathrm{SN}}/a_{\mathrm{LMS}}},
\label{mathcalN}
\end{equation} 

\noindent
we may thus rewrite Eq.~(\ref{NkR1}) such that

\begin{equation}
n_k = _{\scriptstyle \mu \rightarrow \infty}^{~\displaystyle \mathrm{lim}}\frac{1}{k!}\int\limits_{0}^{\mu} e^{-\mu'}\mu'^k \mathcal{\scriptstyle N}(\mu')\mathrm{d}\mu' \equiv \langle \mathcal{\scriptstyle N} \rangle_k,
\label{NkR2}
\end{equation}

\noindent
i.e., $n_k$ is equal to the expectation value of $\mathcal{\scriptstyle N}$. In this respect, $w_{\mathrm{ISM}}$ should be regarded as a weighting function and the subscript $k$ on the bracket indicates the dependence on $k$. An approximate solution to this integral can be found by expanding $\mathcal{\scriptstyle N}(\mu)$ in a Taylor series around $\mu=k$. Including the second-order term, we get

\begin{equation}
n_k \simeq \mathcal{\scriptstyle N}(k)+\mathcal{\scriptstyle N}'(k)+\frac{1}{2}(k+1) \mathcal{\scriptstyle N}''(k),
\label{Nkexp}
\end{equation}

\noindent
where, e.g., $\mathcal{\scriptstyle N}'\equiv \mathrm{d}\mathcal{\scriptstyle N}/\mathrm{d}\mu$. Note that if $\mathcal{\scriptstyle N}(\mu)$ is close to linear, then $n_k\simeq \mathcal{\scriptstyle N}(k)~(+\mathrm{const})$.

\par

There is an interesting limiting case when assuming a static ISM and a constant mixing volume. Let $\overline{V}_{\mathrm{mix}}$ be the typical volume affected by each SN such that the ejected material is instantaneously mixed within this volume. Using Eq.~(\ref{asnpsi}), the expression for $\mu(t)$ in Eq.~(\ref{muoft}) is then reduced to 

\begin{equation}
\mu(t)=a_{\mathrm{SN}}\overline{V}_{\mathrm{mix}}\int\limits_{0}^{t}\psi(t')\mathrm{d}t',
\label{muoftred}
\end{equation}

\noindent
A closed expression for $n_k$ can be found such that 

\begin{eqnarray}
n_{k} & = & _{\scriptstyle t \rightarrow \infty}^{~\displaystyle \mathrm{lim}}\frac{1}{k!}\int\limits_{0}^{t} a_{\mathrm{LMS}}e^{-\mu'} \mu'^k \psi(t')\mathrm{d}t' \nonumber \\
& = & \left[ \begin{array}{ll}
\mu' = a_{\mathrm{SN}}\overline{V}_{\mathrm{mix}}\int\limits_0^{t'}\psi(t'')\mathrm{d}t'' \\
\mathrm{d}\mu' = a_{\mathrm{SN}}\overline{V}_{\mathrm{mix}}\psi(t')\mathrm{d}t'
\end{array} \right] \nonumber \\
& = & \frac{1/\overline{V}_{\mathrm{mix}}}{a_{\mathrm{SN}}/a_{\mathrm{LMS}}} \lim_{\mu\to\infty} \frac{1}{k!}\int\limits_{0}^{\mu} e^{-\mu'} \mu'^k \mathrm{d}\mu'  =  \frac{1/\overline{V}_{\mathrm{mix}}}{a_{\mathrm{SN}}/a_{\mathrm{LMS}}}.
\label{Nkvol}
\end{eqnarray} 

\noindent
The ratio is written in this form to show the connection to $\mathcal{\scriptstyle N}$ in Eq.~(\ref{mathcalN}). This result tells us that there are as many (low-mass) stars enriched by $1$ SN as there are stars enriched by $2$ SNe, by $3$ SNe, and so on; i.e., $n_k$ is independent of $k$, even for a varying SFR! This is to be expected since $u_{\mathrm{SN}}(t)=a_{\mathrm{SN}}\psi(t)$, meaning that mapping the chemical inhomogeneities by low-mass star formation occurs at the same rate as the inhomogeneities are generated. It is a limiting case since any time-dependent mixing volume growing with time would result in a decreasing $n_k$ with increasing $k$. This result was used without proof in Karlsson \& Gustafsson (2001, hereafter KG)\nocite{kg01}.

\par

In their work on the evolution of SN remnants, Cioffi et al.\ (1988)\nocite{cioffi88} find that the final radius of the remnant, when it merges with the ambient ISM, is proportional to $\rho^{-18/49}$. The amount of swept-up mass is thus proportional to $\rho^{-5/49}$. Let us neglect this weak dependence on density and assume that the material from a typical SN becomes diluted within a constant mixing mass. The mixing volume is then given by \mbox{$V_{\mathrm{mix}}(t)=\overline{M}_{\mathrm{mix}}/\rho(t)$,} where $\overline{M}_{\mathrm{mix}}$ is the mixing mass and where $\rho(t)$ is determined by Eq.~(\ref{drhodt}). Again assuming instantaneous mixing within the mixing volume, we have

\begin{equation}
\mu(t)=a_{\mathrm{SN}}\overline{M}_{\mathrm{mix}}\int\limits_{0}^{t}\psi(t')/\rho(t')\mathrm{d}t'. 
\end{equation}

\noindent
Hence, $\mathrm{d}\mu/\mathrm{d}t\propto \psi/\rho$. Using Eqs. (\ref{mathcalN}$-$\ref{Nkexp}) we obtain

\begin{equation}
n_{k} = \frac{1/\overline{M}_{\mathrm{mix}}}{a_{\mathrm{SN}}/a_{\mathrm{LMS}}} \langle \rho \rangle_k \simeq \frac{1/\overline{M}_{\mathrm{mix}}}{a_{\mathrm{SN}}/a_{\mathrm{LMS}}} \rho(k),
\label{Nkmass}
\end{equation} 

\noindent
by keeping only the zeroth order term. This expression differs somewhat from the one in Eq.~(\ref{Nkvol}). Suppose that the SN rate ($\propto \psi$) increases with time. The evolution of the chemical inhomogeneities, characterized by the average number of overlapping mixing volumes, then speeds up. This is compensated for by a faster mapping of the inhomogeneities via the higher SFR ($\psi$). However, if the density of the ISM also increases with time (e.g., due to infall), the size of the mixing volumes decreases and evolution of the inhomogeneities is slowed down accordingly. Since the SFR is unaffected, more stars are therefore able to map this state of chemical inhomogeneity. Thus, the number of stars enriched by $k$ SNe increases with increasing $k$, for an increasing density as indicated by Eq.~(\ref{Nkmass}). In fact, if $\psi \propto \rho$, then $n_k\propto \psi$, contrary to the result in Eq.~(\ref{Nkvol}). 

\par

In the derivation of the two last results, we assumed that the mixing volumes have a certain, pre-determined size and that the SN material is instantaneously mixed within this volume. However, due to the continuous agitation of the ISM the mixing volumes will grow with time. This means, among other things, that there exists a distribution of mixing masses rather than a single constant mixing mass.

\subsection{The distribution of mixing masses} \label{theory_mixingmass}
\noindent
We now address the question of how much an element ejected in an SN explosion is diluted before being locked up in a subsequently formed star. Assuming homogeneous mixing, the amount of dilution is proportional to the mass of the interstellar material inside the mixing volume associated with the SN. For a time-dependent density $\rho(t)$, the total mass $M$ inside this mixing volume not depends only on the size of $V_{\mathrm{mix}}$ but also on when $V_{\mathrm{mix}}$ was created by the SN explosion. Let $\tau_{V}$ be the time that has elapsed since the creation of $V_{\mathrm{mix}}$. Thus, $\tau_{V}$ denotes the age of $V_{\mathrm{mix}}$ and is a measure of its size. If the SN exploded at time $t-\tau_{V}$, mass $M$ is given by the two-dimensional function $g_{M}$, where

\begin{equation}
g_{M}(t,\tau_{V}) = 
\left\{ \begin{array}{ll}
\!\!\int\limits_{t-\tau_{V}}^{t}\!\!\!\dot{V}_{\mathrm{mix}}(t'-t+\tau_{V})\rho(t')\mathrm{d}t', & t\ge \tau_{V} \\
0, & t<\tau_{V}.  \end{array}
\right.
\label{vrmass}
\end{equation} 

\noindent
The mass of enriched gas only grows by mixing of the {\it surface} of $V_{\mathrm{mix}}$ with the surrounding material. The mixing mass can neither change nor, in particular, decrease, due to density changes {\it within} $V_{\mathrm{mix}}$. The term $V_{\mathrm{mix}}\dot{\rho}$ is therefore not included in the integral in Eq. (\ref{vrmass}).

\par

\begin{figure}[t]
 \resizebox{\hsize}{!}{\includegraphics{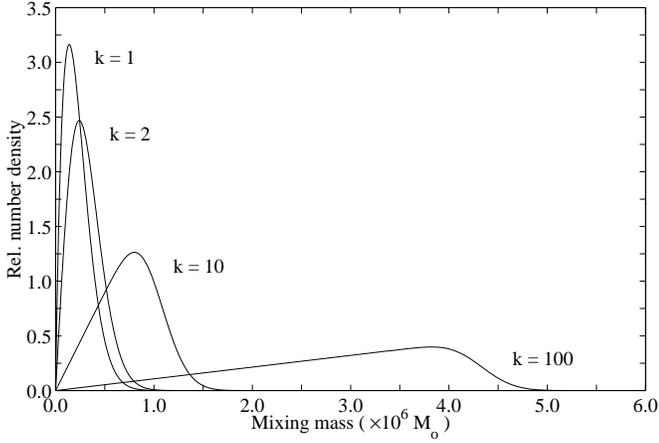}}
 \caption{The distribution of mixing masses in regions enriched by \mbox{$k=1,2,10,$} and $100$ SNe as given by Eq.~(\ref{derivedfm}) for a constant SFR (see Model A in Sect.~\ref{appl_modela}). For example, for $k=10$ all mixing masses $M_i$ in Eqs.~(\ref{n_AH}) and (\ref{n_AB}) should be picked from the density function $f_{M_{k=10}}$.}
 \label{Mmix_fig}
\end{figure}

Now, at time $t\ge \tau_{V}$, mixing volumes of size $V_{\mathrm{mix}}(\tau_{V})$ appear at a rate $u_{\mathrm{SN}}(t-\tau_{V})$, which is a measure of how many mixing volumes there are of this size at time $t$. Thus, the probability that such a mixing volume overlaps a region already enriched by $k-1$ SNe at time $t$ is proportional to $V_{\mathrm{mix}}(\tau_{V}) w_{\mathrm{ISM}}(k-1,t) u_{\mathrm{SN}}(t-\tau_{V})$. The probability of forming a still-existing star in this region in time interval $[t,t+\mathrm{d}t]$ and age interval $[\tau_{V},\tau_{V}+\mathrm{d}\tau_{V}]$ is then proportional to $a_{\mathrm{SN}}a_{\star}(t)V_{\mathrm{mix}}(\tau_{V}) w_{\mathrm{ISM}}(k-1,t)\psi(t-\tau_{V}) \psi(t)\mathrm{d}t\mathrm{d}\tau_{V}$, where we have used Eq.~(\ref{asnpsi}) to write $u_{\mathrm{SN}}$ in terms of $\psi$. Integrating this expression over the region $\Delta D_{M}$ in the $(t,\tau_{V})$-plane gives the total probability $f_{M_k}$ of having a mixing mass of size $M$ in regions enriched by $k$ SNe (cf.\ the treatment in KG). Thus, 

\begin{eqnarray}
f_{M_k} & = & c_k\!\!\!\!\int\limits_{\Delta D_{M}}\!\!\!\!a_{\star}(t)V_{\mathrm{mix}}(\tau_{V}) w_{\mathrm{ISM}}(k-1,t)\times \nonumber\\
 &\times & \psi(t-\tau_{V})\psi(t)\mathrm{d}t\mathrm{d}\tau_{V},
\label{generalfm}
\end{eqnarray}

\noindent
where the SN fraction $a_{\mathrm{SN}}$ has been included in the normalization constant $c_k$. The differential integration region $\Delta D_{M}$ is defined by function $g_{M}(t,\tau_{V})$ in Eq.~(\ref{vrmass}), such that 

\begin{equation}
\Delta D_{M}=\{(t,\tau_{V});~M<g_{M}(t,\tau_{V})\le M+\mathrm{d}M\}.
\end{equation}   

\par 

We may simplify Eq.~(\ref{generalfm}) by assuming that the density of the ISM is constant. The mixing mass $M$ is then directly proportional to the mixing volume such that 

\begin{equation}
M=\overline{\rho} V_{\mathrm{mix}}(\tau_{V})\equiv g_{M}(\tau_{V}),
\label{crmass}
\end{equation}

\noindent
where $\overline{\rho}$ is a typical density. Since the mass is determined by the size of $V_{\mathrm{mix}}$ alone, the integration over $\tau_{V}$ is redundant and Eq.~(\ref{generalfm}) reduces to a one-dimensional integral

\begin{equation}
f_{M_k} = \,\,\,\tilde{c}_kM\!\!\!\!\!\!\!\int\limits_{g_M^{-1}(M)}^{\tau_{\mathrm{G}}}\!\!\!\!\!\! w_{\mathrm{ISM}}(k-1,t)\psi(t-\tau_{V})\psi(t)\mathrm{d}t,
\label{derivedfm}
\end{equation}

\noindent
where we used Eq.~(\ref{crmass}) to write $V_{\mathrm{mix}}(\tau_{V})$ in terms of $M$, which can be taken outside the integral. The factors $a_{\mathrm{LMS}}$ and $1/\overline{\rho}$ have been included in the normalization constant $\tilde{c}_k$. Recall that $\tau_{\mathrm{G}}$ denotes the age of the Galaxy. Since mixing masses of size $M=g_{M}(\tau_{V})$ do not exist at times $t<\tau_{V}$, a lower integration limit is introduced at $t=\tau_{V}$, making it, as indicated, a function of $M$.

\par

As shown in Fig.~\ref{Mmix_fig}, the distribution $f_{M_k}$ is broader and the average mixing mass is larger for higher values of $k$, which is a result of the turbulent diffusion of the ISM, mimicked by the growing mixing volumes. The material ejected by the very first SNe has continuously been diluted and spread over large regions. Their mixing masses are therefore large.

\section{The stochastic chemical enrichment model} \label{stoch} 
\noindent
Let us now develop an inhomogeneous chemical-enrichment model based on the mixing-volume picture of the ISM. Suppose that an SN explodes inside the mixing volume of an earlier SN. At this point the mass fraction of element $A$ in the first mixing volume is $p_{A}(m_1)/M_1$, where $p_{A}(m_1)$ is the yield of $A$ from an SN of mass $m_1$ and $M_1$ is the total mass within the first mixing volume. The total mass of $A$ within the second mixing volume is given by the yield of $A$ ejected from the second SN plus the fraction of $A$ from the first SN, i.e., $p_{A}(m_2)+p_{A}(m_1)M_2/M_1$. The mass fraction is  thus $p_{A}(m_2)/M_2+p_{A}(m_1)/M_1$. Similarly, assuming an initially primordial ISM, the abundance of element $A$ in regions enriched by $k$ SNe is then given by the sum

\begin{equation}
(A/\mathrm{H})=\frac{1/m_A}{X_{\mathrm{H}}/m_{\mathrm{H}}}\times \sum\limits^{k}_{i=1}\frac{p_A(m_i)}{M_i} \equiv y_{k(A)},
\label{n_AH}
\end{equation}

\noindent
where $m_i$ is the progenitor mass of the $i$th SN, and $M_i$ is the mass of the mixing volume belonging to that SN. The atomic masses of element $A$ and hydrogen are denoted by $m_A$ and $m_{\mathrm{H}}$, respectively. In the extremely metal-poor regime the hydrogen mass fraction does not change drastically from its primordial value, and we will assume that $X_{\mathrm{H}}=0.75$ ($Y_{\mathrm{p}}=0.25$, Huey et al.\ 2004\nocite{huey04}). However, for higher metallicities a first-order correction can be introduced such that $X_{\mathrm{H}}=-3.00 Z+0.75$, where $Z$ is the average mass fraction of metals (based on data taken from Cox 2000\nocite{cox00}).  

\par

For the sake of completeness we shall also write down the expression for an abundance ratio of two heavy elements $A$ and $B$. From Eq.~(\ref{n_AH}) we see that

\begin{equation}
(A/B)=\frac{m_{B}}{m_{A}} \times \frac{\sum\limits^{k}_{i=1} p_A(m_i)/M_i}{\sum\limits^{k}_{i=1}p_B(m_i)/M_i} \equiv y_{k(AB)}.
\label{n_AB}
\end{equation} 

\noindent
Comparing this expression with Eq.~(23) in KG we may identify the weights $\xi_i$ that were introduced in an ad hoc way, in order to mimic the effects of large-scale mixing and dilution of the SN material. Eq.~(\ref{n_AB}) tells us that $\xi_i\equiv 1/M_i$. Although it was shown in KG that the chemical abundance patterns in diagrams relating the ratios of two heavier elements (denoted $A/A$ diagrams) are insensitive to the weights, we now have a theory for calculating these weights.

\par

The total metallicity distribution $f_Z$ of low-mass stars, with $Z=(A/\mathrm{H})$, may formally be expressed as a sum of integrals using the formalism in KG by replacing their Eq.~(15) with our new expression for the abundance $y_{k(A)}$. However, $f_Z$ can also be found directly by Monte-Carlo simulations, a method used in this study.  

\par

For each number $k=1,...,n$ we generate $w_k \times 10^6$ low-mass stars, where each star has a surface abundance of $A$ relative to H determined by Eq.~(\ref{n_AH}). The weight \mbox{$w_k=n_k/\sum_{k=1}^{n}n_k$} is the relative number of stars enriched by $k$ SNe, where $n_k$ is given by Eq.~(\ref{Nkexp}). The value $10^6$ is the average number of realizations per number of enriching SNe. Thus, the total number of generated stars is $n\times 10^6$, where the maximum number $n$ of enriching SNe is chosen to be large enough such that $99.9\%$ of all the low-mass stars enriched by $n$ SNe have a metallicity $[A/\mathrm{H}]>-2.5$. The stellar masses $m_i$ are randomly picked from the probability density function $f_m=\phi(m)$, i.e., the adopted IMF, and the mixing masses $M_i$ are picked from the density functions $f_{M_k}$ derived in Eq.~(\ref{generalfm}). Note that for the stars enriched by $k$ SNe, all $k$ mixing masses should be picked from the $k$th density function (see Fig.~\ref{Mmix_fig}). All stars enriched by $k$ SNe with the same metallicity are now binned together to form, after normalization, the partial probability density function $f_{k}(Z)$. Finally, the total probability density function is given by

\begin{equation}
f_Z(Z)=\sum\limits_{k=1}^{n}w_k\times f_{k}(Z),
\label{fZ}
\end{equation}

\noindent
where the contribution of $f_{k}$ to $f_Z$ is given by $w_k$, the relative number of stars enriched by $k$ SNe (cf. Eq.~(22) in KG).

\section{Application of the theory} \label{application}
\noindent
In the present theory the chemical enrichment of a metal-poor system is completely determined by the five parameter functions $V_{\mathrm{mix}}(t),\psi(t),\rho(t),\phi(m)$, and $p_{A}(m)$, whereof three ($V_{\mathrm{mix}}(t),\phi(m),p_{A}(m)$) are more or less universal (although not necessarily known), while the SFR ($\psi(t)$) and the density evolution ($\rho(t)$) are specific for each system. The intention in the following application of the theory is not to present a full parameter study, but merely to investigate the effects caused by changing the SFR and density evolution. First, however, the choice of the other parameters will be commented on.

\begin{center}
\begin{table}[t]
\caption{Input parameters used in application of the theory}
  \label{parameters}
  \begin{tabular}{lll}
     \hline
     \hline
     \\*[-0.5em]
\hfill\footnotesize{Parameter}\hfill{} & \hfill\footnotesize{Symbol}\hfill{} & \hfill\footnotesize{Value}\hfill{} \\
     \\*[-0.5em]
     \hline
     \\*[-0.8em]
\footnotesize{SN fraction$^{\mathrm{a}}$} & \footnotesize{$a_{\mathrm{SN}}$} & \footnotesize{$0.002$} \\*[0.1em]
\footnotesize{Low-mass star fraction$^{\mathrm{a,b}}$} & \footnotesize{$a_{\mathrm{LMS}}$} & \footnotesize{$0.835$} \\*[0.1em]
\footnotesize{Typical density (Halo)$^{\mathrm{c}}$} & \footnotesize{$\overline{\rho}$} & \footnotesize{$2.05\!\times\!10^{-25}$ g cm$^{-3}$} \\*[0.1em]
\footnotesize{Density normalization} & \footnotesize{$\rho_0$} & \footnotesize{$7.18\!\times\!10^{-25}$ g cm$^{-3}$} \\*[0.1em]
\footnotesize{Average SN rate (Halo)$^{\mathrm{d}}$} & \footnotesize{$\overline{u}$} & \footnotesize{$0.25$ kpc$^{-3}$ Myr$^{-1}$} \\*[0.1em]
\footnotesize{SN rate normalization} & \footnotesize{$u_0$} & \footnotesize{$1.0$ kpc$^{-3}$ Myr$^{-1}$} \\*[0.1em]
\footnotesize{Expansion rate$^{\mathrm{e}}$} & \footnotesize{$\sigma_{\mathrm{mix}}$} & \footnotesize{$7\!\times\! 10^{-4}$ kpc$^2$ Myr$^{-1}$} \\   
     \\*[-0.8em]
     \hline
     \\*[-0.8em]
  \end{tabular}

\hspace*{5pt}\footnotesize{$^{\mathrm{a}}$IMF taken from Kroupa et al.\ (1991)}\nocite{ketal91}  \\*[0.05em]
\hspace*{5pt}\footnotesize{\indent$^{\mathrm{b}}$Stellar lifetimes taken from Portinari et al.\ (1998)\nocite{petal98}} \\*[0.05em]
\hspace*{5pt}\footnotesize{\indent$^{\mathrm{c}}$Corresponds to a particle density of $0.1$  cm$^{-3}$} \\*[0.05em]
\hspace*{5pt}\footnotesize{\indent$^{\mathrm{d}}$Estimated from Fig.~\ref{usn_fig}} \\*[0.05em]
\hspace*{5pt}\footnotesize{\indent$^{\mathrm{e}}$Expansion rate estimated from Bateman \& Larson (1993)\nocite{bl93}} \\
\end{table}
\end{center}

\begin{figure}[t]
 \resizebox{\hsize}{!}{\includegraphics{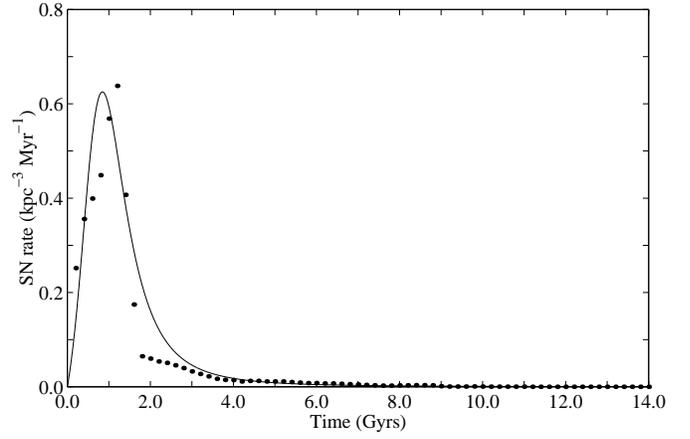}}
 \caption{The rate of Type~II SNe in the Galactic halo as a function of time. The dots are the result of a chemodynamical simulation by Samland et al.\ (1997), with SN rates given in their Fig.~8. A small correction was made for the number of SNe exploding in the pre-disk while actually belonging to the halo. The curve, given by Eq.~(\ref{usn}), is the adopted SN rate in Models B and C.}
 \label{usn_fig}
\end{figure}

\subsection{The mixing volume growth, $V_{\mathrm{mix}}(t)$} \label{appl_vmix}
\noindent
The different evolutionary phases of an SN remnant are well understood; and in the ideal case of a blast wave expanding into a homogeneous medium assuming no radiative cooling of the interior, the radius $R_{\mathrm{SNR}}$ of the remnant is described by power-law solutions of the form $R_{\mathrm{SNR}}\propto t^{\eta}$, where the value $\eta$ depends on the phase the remnant is in (Taylor 1950\nocite{taylor50}; Sedov 1959\nocite{sedov}; McKee \& Ostriker 1977\nocite{mcko77}). However, this initial spread of the SN material is confined to a small volume and the chance of immediately forming stars in the swept-up gas is probably not very large. The successive, large-scale mixing of the cooled SN material---caused, e.g., by turbulent motions---must therefore be included. Bateman \& Larson (1993)\nocite{bl93} estimate the dispersal of iron, ascribing the motions in the ISM to different random-walk processes where the rms distance $R_{\mathrm{rms}}$ increases as $t^{1/2}$. If so, the mixing volume should on average increase as

\begin{equation}
V_{\mathrm{mix}}(t)=\frac{4\pi}{3}R_{\mathrm{rms}}^3=\frac{4\pi}{3}\left(\sigma_{\mathrm{mix}}t \right)^{3/2}~~~\mathrm{kpc}^3,
\label{vmix}
\end{equation}

\noindent
where $\sigma_{\mathrm{mix}}=\frac{2}{3}\sum_i v_i^2\tau_i$ is a combined ``cross-sectional expansion rate'' for the processes responsible for the bulk distribution of heavy elements. These are expected to be cloud motions and turbulence in the diffuse medium, assuming the spreading in the diluted hot ionized medium can be neglected (see Bateman \& Larson 1993\nocite{bl93}). The $v_i$ in the sum denotes the mean velocity in each phase of the gas, and $\tau_i$ is the time between scattering events. From the data given in Bateman \& Larson (1993\nocite{bl93}) we find an overall expansion rate of \mbox{$\sigma_{\mathrm{mix}}=7\times 10^{-4}$ kpc$^2$ Myr$^{-1}$} (see Table~\ref{parameters}). Neglecting the relatively small initial expansion of the SN remnant itself, an approximate estimate of the total mixing volume growth is given by Eq.~(\ref{vmix}), where $t$, measured in Myr, denotes the time after the SN explosion.

\subsection{The IMF, $\phi(m)$} \label{appl_imf}
\noindent
We adopt the IMF suggested by Pagel (1997\nocite{pagel97}). It is a composite of the IMF determined by Kroupa et al.\ (1991)\nocite{ketal91} for stars below $m=1.0$ $\mathcal{M_{\odot}}$ and two power-law segments approximating the IMF by Scalo (1986\nocite{scalo86}) for stars of higher mass. With the normalization given by Eq.~(\ref{imf}) we have

\begin{equation}
\phi(m) = 
\left\{ \begin{array}{llll}
0.506m^{-0.85}, & 0.10\le m \le 0.50 \\
0.253m^{-1.85}, & 0.50\le m \le 1.00 \\
0.253m^{-3.40}, & 1.00\le m \le 3.16 \\
0.113m^{-2.70}, & 3.16\le m \le 100. \end{array}
\right.
\label{imfpagel}
\end{equation} 

\noindent
For this IMF, the fraction of stars that are massive enough \mbox{($>8~\mathcal{M_{\odot}}$)} to eventually explode as core-collapse SNe is \mbox{$a_{\mathrm{SN}}=0.002$} (see Table~\ref{parameters}). Moreover, using the table \mbox{($Z=0.0004$)} of stellar lifetimes by Portinari et al.\ (1998)\nocite{petal98} we estimate that $\langle a_{\star}\rangle=a_{\mathrm{LMS}}=0.835$, averaged over the first billion years.

\subsection{The stellar yield, $p_{A}(m)$} \label{appl_yield}
\noindent 
Many observational studies use iron (Fe) as a reference element. In order to make a direct comparison easier we shall therefore use $[\mathrm{Fe}/\mathrm{H}]$ as the metallicity indicator. We adopt the Fe yields for Population III core-collapse SNe calculated by Umeda \& Nomoto (2002\nocite{un02}), assuming that all the SNe in the mass interval $13-30~\mathcal{M_{\odot}}$ have an explosion energy of $E=10^{51}$ erg. The fact that the Fe yield is uncertain due to the unknown amount of fall-back of $^{56}$Ni onto the neutron star will not alter the general conclusions drawn in this paper.

\par

At later epochs, Type~Ia SNe become an additional source of Fe, which is not taken into account in the current model. However, observations have shown that up to $[\mathrm{Fe}/\mathrm{H}]\simeq -1$ the ISM retained a constant $[\alpha/\mathrm{Fe}]$, the $\alpha$-elements being Mg, Si, Ca, and possibly Ti, suggesting that Type~II SNe are the major source of Fe. Kobayashi et al.\ (1998\nocite{kobayashi98}) point out that a metallicity-dependent rate of Type~Ia SNe could explain this relatively late decrease in $[\alpha/\mathrm{Fe}]$. In Sect. \ref{sectsnia}, the model is extended to incorporate Fe enrichment by Type~Ia SNe. The level of such contamination is found to be small for extremely metal-poor stars.

\begin{center}
\begin{table}[t]
\caption{Integration times}
  \label{inttimes}
  \begin{tabular}{lllll}
     \hline
     \hline
     \\*[-0.5em]
\hfill\footnotesize{ }\hfill{} & \hfill\footnotesize{Model A}\hfill{} & \hfill\footnotesize{Model B}\hfill{} & \hfill\footnotesize{Model C}\hfill{} &\hfill\footnotesize{ }\hfill{} \\
     \\*[-0.5em]
     \hline
     \\*[-0.8em]
\footnotesize{$n^{\mathrm{a}}$} & \footnotesize{$240$} & \footnotesize{$135$} & \footnotesize{$275$} & \\*[0.1em]
\footnotesize{${\mu_{u}}^{\mathrm{b}}$} & \footnotesize{$326.4$} & \footnotesize{$200.1$} & \footnotesize{$367.4$} & \\*[0.1em]
\footnotesize{${t_{u}}$ Myr$^{\mathrm{c}}$} & \footnotesize{$1121$} & \footnotesize{$924$} & \footnotesize{$1117$} & \\
     \\*[-0.8em]
     \hline
     \\*[-0.8em]
  \end{tabular}

\hspace*{5pt}\footnotesize{$^{\mathrm{a}}$Maximum number of enriching SNe (see Sect.~\ref{stoch})} \\*[0.05em]
\hspace*{5pt}\footnotesize{$^{\mathrm{b}}$Required $\mu$ for $k=n$ SNe, according to Eq.~(\ref{reqmu}) }\\*[0.05em]
\hspace*{5pt}\footnotesize{$^{\mathrm{c}}$Effective upper integration limit for $k=n$ SNe} \\

\end{table}
\end{center}

\subsection{The models} \label{appl_models}
\noindent
Three models are specified by different sets of the parameters $\psi(t)$ and $\rho(t)$. These parameters will be prescribed, i.e., they will not be explicitly calculated from expressions such as Eq. (\ref{drhodt}). The parameters $V_{\mathrm{mix}}(t),~\phi(m)$, and $p_{\mathrm{Fe}}(m)$ are the same for all three models.

\begin{figure}[t]
 \resizebox{\hsize}{!}{\includegraphics{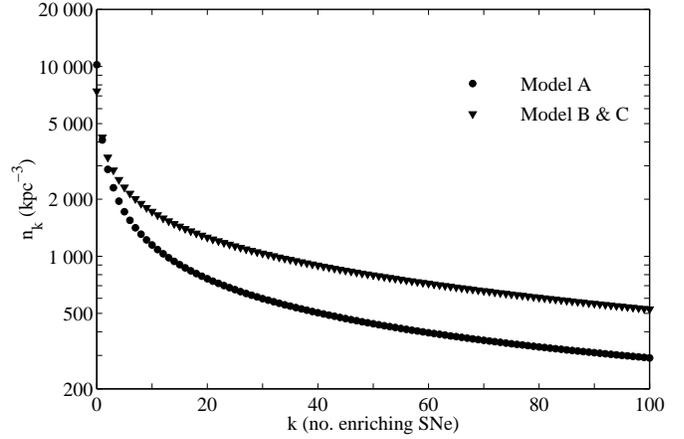}}
 \caption{The number density $n_k$ of low-mass Galactic halo stars as a function of the number $k$ of enriching SNe. The number density of stars decreases with increasing $k$, for Model A (filled circles), as well as for Model B \& C (filled triangles). This effect is caused by the fact that the mixing volumes are allowed to grow with time.}
 \label{nk_fig}
\end{figure}

\subsubsection{Model A} \label{appl_modela}
\noindent
Model A corresponds to the extreme inflow model by Larson (1972\nocite{larson72}), which is characterized by a constant SFR/SN rate $\overline{u}$ and a constant gas density $\overline{\rho}$ (see Table~\ref{parameters}). This means that the accretion compensates exactly for the loss of gas in star formation. 

\par

The past SFR for the Galactic halo is not very well known. Assuming that the rate of SNe in the early Galaxy is comparable to the present rate of $1$ SN every $70$ years, we obtain a rate of $u_{\mathrm{SN}}=0.05$ kpc$^{-3}$ Myr$^{-1}$, averaged over the spheroid of $40$ kpc radius. This is probably a conservative estimate. In their chemodynamical simulation of the Galaxy, Samland et al.\ (1997\nocite{setal97}) deduced the rate of Type~II SNe in the Galactic halo as a function of time (see Fig.~\ref{usn_fig}). Using their data, we infer an SN rate of $u_{\mathrm{SN}}=0.25$ kpc$^{-3}$ Myr$^{-1}$, averaged over the first billion years. This is the adopted rate. Moreover, the average particle density is taken to be $0.1$ cm$^{-3}$, which should be a reasonable value for the interstellar gas density of the inner Galactic halo, where most of the stars presumably were formed.

\subsubsection{Model B} \label{appl_modelb}
\noindent
Model B is an extension of Model A with an SFR that is allowed to vary with time. Recalling that $\psi(t)=u_{\mathrm{SN}}(t)/a_{\mathrm{SN}}$, we shall assume an SN rate of the form 

\begin{equation}
u_{\mathrm{SN}}(t)=u_0\frac{(t/1000)}{1+(t/1000)^4}~~~\mathrm{kpc}^{-3}~\mathrm{Myr}^{-1},
\label{usn}
\end{equation}

\noindent
where $t$ is in Myr and the normalization $u_0$ is given in Table~\ref{parameters}. The choice of function is arbitrary and is only used to demonstrate the effect of a time-dependent SFR. However, as shown in Fig.~\ref{usn_fig}, this function resembles the evolution of the SN rate given by Samland et al.\ (1997\nocite{setal97}), with a sharp rise at early times followed by a slow decline. As in Model A, the density is given by $\rho=\overline{\rho}$.

\subsubsection{Model C} \label{appl_modelc}
\noindent
In order to investigate the effects of a time-dependent density we assume that $\rho\propto \psi$. Thus, $\rho(t)$ has the same form as the expression for $u_{\mathrm{SN}}(t)$ in Eq.~(\ref{usn}), with a normalization $\rho_0=7.18\times 10^{-25}$ g cm $^{-3}$, which corresponds to a maximum particle density of $0.2$ cm$^{-3}$ at $t=760$ Myr. As in Model B, the SN rate is given by Eq.~(\ref{usn}).

\section{Results and discussion} \label{results}
\noindent
\subsection{Integration times} \label{results_times}
\noindent
Since the probability density function $w_{\mathrm{ISM}}$ decreases very fast shortly after it has reached its maximum, the expressions for $n_k$ and $f_{M_k}$ do not have to be integrated up to $t=\tau_{\mathrm{G}}$. An effective upper integration limit $t_u$ may be introduced. This corresponds to an upper limit on $\mu$ such that $\mu_u=\mu(t_u)$. The expression for $\mu_u$ is chosen to be

\begin{equation}
\mu_u=\langle\mu \rangle+\frac{11}{2}\sigma=\frac{11}{2}\sqrt{n+1}+n+1,
\label{reqmu}
\end{equation}

\noindent
where $\langle\mu \rangle=n+1$ is the mean and $\sigma=\sqrt{n+1}$ is the standard deviation of $w_{\mathrm{ISM}}$ for $k=n$, i.e., the maximum number of SNe. With a factor $11/2$, the integral of $w_{\mathrm{ISM}}$ up to $\mu_u$ for $n=100$ differs only one part in a million from the total integral ($\equiv 1$). Table~\ref{inttimes} shows the upper integration limits for all three models. Evidently, the population of extremely metal-poor stars with a metallicity $[\mathrm{Fe}/\mathrm{H}]\le -2.5$ is formed after $\sim 1$ Gyr, irrespective of the choice of model. 

\begin{figure}[t]
 \resizebox{\hsize}{!}{\includegraphics{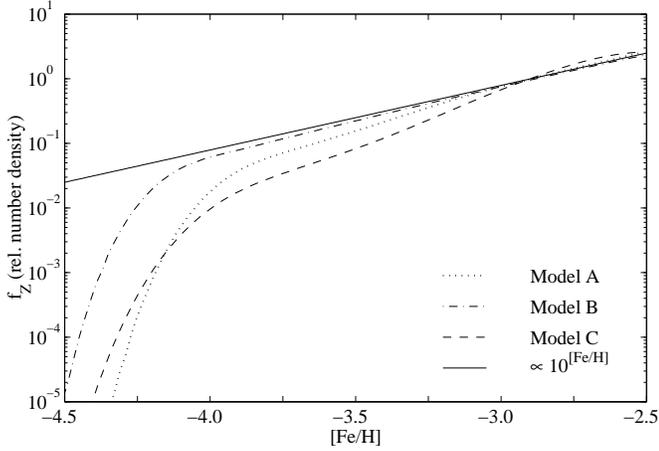}}
 \caption{Relative number densities $f_Z$ of stars as a function of metallicity, measured by [Fe/H], for different parameters $\psi$ and $\rho$. See Sect.~\ref{appl_models} for details. The Fe yields are taken from Umeda \& Nomoto (2002).}
 \label{nZmodels}
\end{figure}

\subsection{The metallicity distribution} \label{results_distr}
\noindent
Before discussing the metallicity distributions we comment briefly on the $k$-dependence of $n_k$, the number density of (still-existing) low-mass stars. As shown in Fig.~\ref{nk_fig}, $n_k$ decreases with increasing number of enriching SNe. This decrease is present in all models. In fact, unless the mixing-volume growth depends on the gas density of the ISM ($\sigma_{\mathrm{mix}}$ is assumed to be constant in the present study), $n_k$ will always decrease with increasing $k$. This result can be ascribed to the properties of $\mathcal{\scriptstyle{N}}$, defined in Eq.~(\ref{mathcalN}), which is a decreasing function of $\mu$ (and therefore time).      

\par   

\begin{center}
\begin{table}[t]
\caption{Results}
  \label{restable}
  \begin{tabular}{llll}
     \hline
     \hline
     \\*[-0.5em]
\hfill\footnotesize{}\hfill{} & \hfill\footnotesize{Model A}\hfill{} & \hfill\footnotesize{Model B}\hfill{} & \hfill\footnotesize{Model C}\hfill{}\\
     \\*[-0.5em]
     \hline
     \\*[-0.8em]
\footnotesize{$n_{\mathrm{III}}~\mathrm{kpc}^{-3}$} & \footnotesize{$10~250$} & \footnotesize{$7~420$} & \footnotesize{$7~420$} \\*[0.1em]
\footnotesize{$a_{\mathrm{III}}$} & \footnotesize{$2.0\!\times\!10^{-2}$} & \footnotesize{$1.4\!\times\!10^{-2}$} & \footnotesize{$1.0\!\times\!10^{-2}$} \\*[0.1em]
\footnotesize{$N_{Z}^{\mathrm{noIII}}(-4)^{\mathrm{a}}~\mathrm{kpc}^{-3}$} & \footnotesize{$98$} & \footnotesize{$772$} & \footnotesize{$105$} \\*[0.1em]
\footnotesize{${a_{Z\le -4}^{\mathrm{noIII}}}^{\mathrm{a}}$} & \footnotesize{$2.2\!\times\!10^{-4}$} & \footnotesize{$1.5\!\times\!10^{-3}$} &  \footnotesize{$1.5\!\times\!10^{-4}$}\\*[0.1em]  
\footnotesize{$F_{1}(-2.5)^{\mathrm{b}}$} & \footnotesize{$0.987$} & \footnotesize{$0.996$} & \footnotesize{$0.958$} \\*[0.1em]
\footnotesize{$k_{d}$ at $[\mathrm{Fe}/\mathrm{H}]=-2.5^{\mathrm{c}}$} & \footnotesize{$142$} & \footnotesize{$74$} & \footnotesize{$158$} \\
     \\*[-0.8em]
     \hline
     \\*[-0.8em]
  \end{tabular}
\hspace*{5pt}\footnotesize{$^{\mathrm{a}}$Number/fraction of stars below $[\mathrm{Fe}/\mathrm{H}]=-4$}, excluding Pop. III\\*[0.05em]
\hspace*{5pt}\footnotesize{$^{\mathrm{b}}$Fraction of stars enriched by $1$ SN below $[\mathrm{Fe}/\mathrm{H}]=-2.5$}\\*[0.05em]
\hspace*{5pt}\footnotesize{$^{\mathrm{c}}$$f_{k_d}$ is the dominant contributor to $f_Z$ at $[\mathrm{Fe}/\mathrm{H}]=-2.5$} \\*[0.05em]

\end{table}
\end{center}

\begin{figure}[t]
 \resizebox{\hsize}{!}{\includegraphics{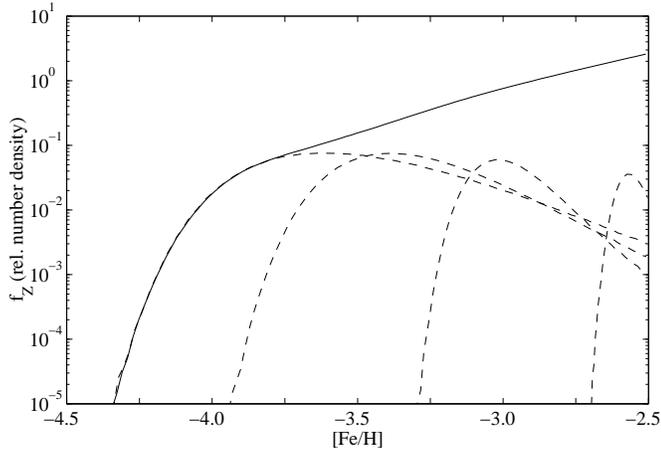}}
 \caption{The number density of extremely metal-poor stars in the Galactic halo as predicted by Model A. The full line is the total number density $f_Z$, while the dashed lines are the partial number densities $f_{k}$ for $k=1,2,10$, and $100$ enriching SNe (cf. Fig.~\ref{Mmix_fig}) Note the long tail of stars enriched by one SN extending beyond $[\mathrm{Fe}/\mathrm{H}]=-2.5$.}
 \label{nZ_fig}
\end{figure}

\begin{figure}[t]
 \resizebox{\hsize}{!}{\includegraphics{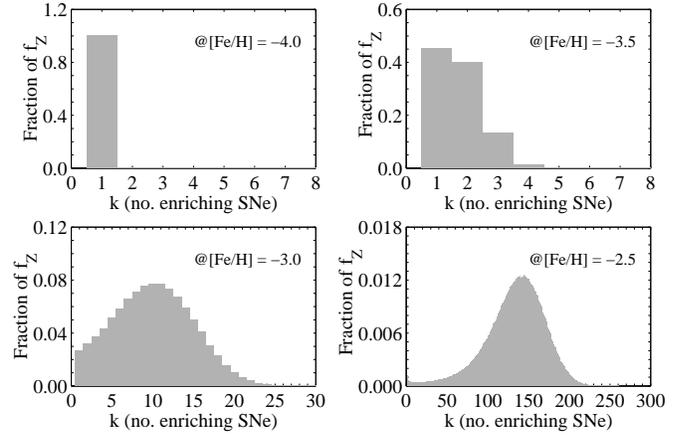}}
 \caption{The fraction of stars enriched by $k$ SNe at metallicity $[\mathrm{Fe}/\mathrm{H}]=-4.0,-3.5,-3.0,$ and $-2.5$ as predicted by Model A. Note the broad distribution at $[\mathrm{Fe}/\mathrm{H}]=-2.5$, peaking at $k_d=142$ (see also Table~\ref{restable}).}
 \label{kdistr_fig}
\end{figure}

Our resulting metallicity distributions for the three models are shown in Fig.~\ref{nZmodels} with [Fe$/$H] as metallicity indicator. The number density of stars drops drastically below [Fe$/$H]$\simeq-4$. This is the low-metallicity tail of stars enriched by a single SN. At [Fe$/$H]$\gtrsim -4$, the number density of stars in Models A and B approach an exponential increase $\propto 10^{[\mathrm{Fe}/\mathrm{H}]}$. Due to an initially increasing density (i.e., infall), Model C shows a suppressed metal-poor end, similar to the effect caused by infall in homogeneous models. At [Fe$/$H]$=-2.5$, the dominant contributing partial density function $f_{k_d}$ consists of stars enriched by as many as $k_d=74 - 158$ SNe (see Table~\ref{restable}). This is in accordance with the estimate by Nissen et al.\ (1994)\nocite{netal94}, who concluded that metal-poor Galactic halo stars must have been enriched by at least $20$ SNe due to the relatively small observed star-to-star scatter in the abundance ratios. The recent work by Carretta et al.\ (2002)\nocite{carretta02} and Cayrel et al.\ (2004)\nocite{cayrel04} verifies a small scatter in many abundance ratios down to $[\mathrm{Fe}/\mathrm{H}]=-4$. As these authors point out, this suggests that even the most metal-poor stars observed today may be enriched by more than one SN. Figs.~\ref{nZ_fig} and \ref{kdistr_fig} show that the model stars are indeed enriched by several SNe, at least down to $[\mathrm{Fe}/\mathrm{H}]=-3.5$ (cf. Argast et al.\ 2000\nocite{argast00}). This result contradicts the conclusion that most stars with $[\mathrm{Fe}/\mathrm{H}]\le-2.5$ have been enriched by individual SNe (e.g., Shigeyama \& Tsujimoto 1998\nocite{st98}). In fact, we cannot be sure that a star has been enriched by a single SN unless we observe it below the cut-off (see e.g., Fig.~\ref{nZ_fig}), which in these models is located at [Fe$/$H]$\simeq-4$. Note however, that the metallicity distributions $f_k$ for low $k$ are very broad and about one percent of all the stars enriched by a single SN have presumably a metallicity $[\mathrm{Fe}/\mathrm{H}]>-2.5$. This fraction is given by $1-F_1(-2.5)$, see Table~\ref{restable}.

\par

Admittedly, in the most metal-poor regime the averaging of a few contributing SNe cannot be enough to explain the small star-to-star scatter, a scatter that also seems to be relatively constant over a wide range in metallicity. However, we shall argue (Paper~II\nocite{kg04}) that a small scatter in diagrams relating a ratio of two heavier elements to the ratio of an element relative to hydrogen (called $A/\mathrm{H}$ diagrams in KG) may be a natural consequence of selection effects favouring contributions from SNe in a specific mass range. 

\par

The simple assumption that the ejected material from $k$ SNe mixes within a constant mixing mass (e.g., KG; Fields et al.\ 2002\nocite{fields02}), implying that $\mathrm{Fe}/\mathrm{H}\propto k$, will underestimate the number of enriching SNe at a given metallicity. For example, an increase in metallicity by $1.5$ dex, from $[\mathrm{Fe}/\mathrm{H}]=-4$ to $[\mathrm{Fe}/\mathrm{H}]=-2.5$, would require a number of $k=30$ SNe. This is a factor of $2$ to $5$ less than predicted by the models presented here. The reason for this is twofold. Both large-scale mixing of the ISM and infall of pristine gas continuously dilute the SN material. Thus, in order to reach a certain metallicity in a specific region in the ISM, a larger number of SNe, on average, have to enrich this region. Note, however, that the distribution is very broad, as is evident from Fig.~\ref{kdistr_fig}.

\subsection{The fraction of Population III stars} \label{results_popiii}
\noindent
Inserting $w_{\mathrm{ISM}}(0,t)$ in the expression for $n_k$ we can calculate the absolute number density of Population III stars, denoted by $n_{k=0}\equiv n_{\mathrm{III}}$. As shown in Table~\ref{restable}, $n_{\mathrm{III}}=7420$ kpc$^{-3}$ for both Models B and C. The higher value for Model A reflects a higher average SFR.

\par

Now, in order to compute the fraction of metal-free stars, we define the total number density of stars $N_{Z}$ with metallicity $\le Z$ as

\begin{equation}
N_{Z}(Z)=\sum\limits_{k=0}^{\infty} F_{k}(Z)n_k,
\label{nzlim}
\end{equation}

\noindent
where $F_{k}$ denotes the distribution function of $f_{k}$ and describes the fraction of stars enriched by $k$ SNe with a metallicity $\le Z$. It is defined by the integral

\begin{equation}
F_{k}(Z)=\int\limits_{-\infty}^{Z} f_{k}(Z')\mathrm{d}Z'.
\label{FkZ}
\end{equation}

\noindent
The fraction of metal-free stars is then formally given by

\begin{equation}
a_{\mathrm{III}}= \frac{n_{\mathrm{III}}}{N_{Z}(\infty)}.
\label{fiii}
\end{equation}

\noindent
Since the integration is terminated at $[\mathrm{Fe}/\mathrm{H}]=-2.5$ in the current application, $N_{Z}(\infty)$ cannot be calculated directly through Eq.~(\ref{nzlim}). However, making use of the observational finding by Carney et al.\ (1996\nocite{carney96}) that $16\%$ of all the stars have a metallicity $[\mathrm{Fe}/\mathrm{H}]<-2.5$, we have $N_{Z}(\infty)=N_{Z}(-2.5)/0.16$. Thus, $a_{\mathrm{III}}= 0.16n_{\mathrm{III}}/N_{Z}(-2.5)$. Using this expression, the fraction of metal-free stars is estimated to $(1-2)\times 10^{-2}$, depending on the model (see Table~\ref{restable}). This is close to the prediction by Oey (2003\nocite{oey03}), who found a value of $a_{\mathrm{III}}\sim (3-4)\times 10^{-2}$, including the effects of additional mixing beyond the local superbubbles. This result thus confirms the discrepancy of almost two orders of magnitude between the observed and the expected number of metal-free stars (Oey 2003\nocite{oey03}).      

\par

If stellar atmospheres are continuously exposed to accretion of enriched material (e.g., Yoshii et al.\ 1981\nocite{yoshii81}), this could, at least qualitatively, explain the observed discrepancy. A truly metal-free star would then be hidden from us by its polluted atmosphere and convection zone. Another possible explanation is that low-mass Population III stars were never formed in the first place (Bond 1981\nocite{bond81}). It has been suggested (e.g., Bromm et al.\ 1999\nocite{bromm99}; Abel et al.\ 2002\nocite{abel02}) that the lack of effective cooling agents (metals) may inhibit low-mass star formation in the primordial ISM. If so, the fraction of stars below, say, $[\mathrm{Fe}/\mathrm{H}]=-4$ should be given by

\begin{equation}
a_{Z\le -4}^{\mathrm{noIII}}= 0.16\frac{N_{Z}(-4)-n_{\mathrm{III}}}{N_{Z}(-2.5)-n_{\mathrm{III}}},
\label{fz4}
\end{equation}

\noindent
again making use of the observational result by Carney et al.\ (1996)\nocite{carney96}. These fractions that exclude Population~III stars (see Table~\ref{restable}) lie in the interval $10^{-4}-10^{-3}$ and are indeed close to the present observational value of $a_{Z\le -4}^{\mathrm{obs}}=4\times 10^{-4}$ (cf. Oey 2003\nocite{oey03}), which reflects the discovery for HE 0107-5240, the most iron-poor star known to date (Christlieb et al.\ 2002\nocite{cetal02}). The exact value of $a_{Z\le -4}^{\mathrm{obs}}$ is as yet, however, uncertain due to small-number statistics.

\subsection{The low-metallicity cut-off} \label{results_cutoff}
\noindent
There has been discussion in the past as to whether the observed cut-off in the number of stars at $[\mathrm{Fe}/\mathrm{H}]\sim -4$ (e.g., McWilliam et al.\ 1995\nocite{mcw95}) is real or whether we should expect a continuous distribution down to $Z=0$. If there is a cut-off, where does it come from? Several authors have argued (e.g., Ryan et al.\ 1996\nocite{rnb96}; Shigeyama \& Tsujimoto 1998\nocite{st98}) that a typical mixing mass of $M\simeq 5\times 10^4~\mathcal{M_{\odot}}$ should be expected, based upon the size of SN remnants when they merge with the ambient ISM (Cioffi et al.\ 1988\nocite{cioffi88}). This results in a limiting metallicity of $[\mathrm{Fe}/\mathrm{H}]\simeq -2.7$, which is at least one order of magnitude greater than the limit inferred from observations. In fact, this should be regarded as an upper limit (for stars enriched by a single SN), since additional mixing, e.g. due to turbulent motions in the ISM, would lower the metallicity. On the other hand, Oey (2003\nocite{oey03}) estimates a lower limit in the range $-4.6\lesssim [\mathrm{Fe}/\mathrm{H}]\lesssim -3.6$ for the largest superbubbles generated by SNe (Oey \& Clarke 1997\nocite{oc97}); and Audouze \& Silk (1995\nocite{as95}) predict a similar threshold at $[\mathrm{Fe}/\mathrm{H}]\sim -4$, assuming that the first generation of $\sim 10^4$ SNe enriched the ISM during a short period of time of $\sim 2$ Myr. 

\par

The statistical approach presented here offers an alternative view of the problem. If we allow large-scale mixing of the ISM and assume reasonable values for the density and SFR of the Galactic halo, the distribution of mixing masses for $k=1$ is relatively peaked and lies between $10^5-10^6~\mathcal{M_{\odot}}$ (see Fig.~\ref{Mmix_fig}). As shown in Fig.~\ref{nZ_fig}, this results in a decline in the number of stars below $[\mathrm{Fe}/\mathrm{H}]=-4$. The definitive value of this threshold is controlled by the star formation process, which we take to be stochastic in nature. Stars enriched by a single SN are formed during a finite period of time. During this time they trace chemical inhomogeneities of specific sizes, corresponding to a specific distribution of mixing masses. In fact, contrary to the claim that the cut-off is determined by the universal property of SN physics (e.g., Ryan et al.\ 1996\nocite{rnb96}), the present approach predicts that a system with, e.g., a different star formation history or density evolution produces a different metallicity distribution with a cut-off determined by the density function $f_{M_{k=1}}$, the probability of having a mixing mass of size $M$ in regions enriched by a single SN.

\subsection{Thermonuclear SNe as an additional Fe source}\label{sectsnia}
\noindent
The current model primarily describes the chemical enrichment of the ISM by the spreading of primary elements produced and ejected in core-collapse (Type~II) SN explosions. Other sources, enriching individual stars in elements with origins different from core-collapse SNe, e.g., intermediate-mass stars, presumably do not have a large impact on the heavy element content in the oldest, extremely metal-poor stars. Intermediate-mass stars, for example, do not synthesize elements heavier than magnesium (excluding the neutron-capture elements). An interesting question is, however, whether the large amounts of $^{56}$Fe ejected in thermonuclear (Type~Ia) SN explosions could have an effect on the Fe abundance in extremely metal-poor stars, presuming that the rate of Type~Ia SNe is high enough at these early epochs. In order to include the chemical enrichment of the Type~Ia SNe, the current model has to be modified slightly.

\subsubsection{The rate of thermonuclear SNe}
\noindent
In order to estimate the largest possible contribution of Type~Ia SNe to the Fe enrichment in extremely metal-poor stars, we adopt a short formation time-scale scenario for the Type~Ia SNe. Following Greggio \& Renzini (1983\nocite{gr83}) the rate $u_{\mathrm{SNIa}}$ of Type~Ia SNe can be estimated by assuming that the SN precursors are mass-accreting C/O white dwarfs in close binary systems. The formation time-scale of a Type~Ia SN is, in this scenario, only determined by the initial mass of the companion star. The delay between the onset of star formation and the onset of the first Type~Ia SNe is therefore merely $\sim 40$ Myr, corresponding to the lifetime of an $8~\mathcal{M_{\odot}}$ star.

\par

Assuming that the Galactic SFR follows an exponential decay, $u_{\mathrm{SNIa}}$ can be properly normalized by requiring that \mbox{$u_{\mathrm{SNIa}}=u_{\mathrm{SNII}}$} at the present epoch (Greggio \& Renzini 1983\nocite{gr83}), where $u_{\mathrm{SNII}}$ is the rate of Type~II SNe. This means that for, e.g., a constant SFR, \mbox{$u_{\mathrm{SNIa}}\simeq 0.8\times u_{\mathrm{SNII}}$} when extrapolated to \mbox{$t=\tau_{\mathrm{G}}$}. The total SN rate replacing Eq. (\ref{asnpsi}) is given by \mbox{$u_{\mathrm{SN}}=u_{\mathrm{SNII}}+u_{\mathrm{SNIa}}$}, where \mbox{$u_{\mathrm{SNII}}=a_{\mathrm{SN}}\psi$}. In the discussion below, Model A will be used as input the model. However, with the inclusion of the Type~Ia SNe, the model will be referred to as Model A$^{\prime}$.

\subsubsection{The fraction of thermonuclear SNe}
\noindent
A low-mass star formed at time $t$ has, on average, been enriched by a total number of $\mu(t)$ SNe out of which $\mu_{\mathrm{SNIa}}(t)$ are SNe of Type~Ia, where $\mu_{\mathrm{SNIa}}$ is given by Eq. (\ref{muoft}) with \mbox{$u_{\mathrm{SN}}=u_{\mathrm{SNIa}}$} (as for $\mu_{\mathrm{SNII}}$). Knowing the probability $f_{\mathrm{SNIa}}(k',k)$ that a star, enriched by a total number of $k$ SNe, is enriched by $k'$ Type~Ia SNe (i.e., \mbox{$k-k'$} Type~II SNe), enables us to determine the total contribution from the Type~Ia SNe in detail. The probability of finding a region enriched by $k'$ Type~Ia SNe and \mbox{$k-k'$} Type~II SNe, respectively, at time $t$ is given by Eq. (\ref{poisson}) such that

\begin{equation}
w_{\mathrm{SNIa}}(k',t)=e^{-\mu_{\mathrm{SNIa}}(t)}\mu_{\mathrm{SNIa}}(t)^{k'}/k'!
\label{wsnia}
\end{equation}

\noindent
and 

\begin{equation}
w_{\mathrm{SNII}}(k-k',t)=e^{-\mu_{\mathrm{SNII}}(t)}\mu_{\mathrm{SNII}}(t)^{k-k'}/(k-k')!
\label{wsnii}
\end{equation}

\noindent
The probability $f_{\mathrm{SNIa}}$ is then determined by the integral

\begin{eqnarray}
f_{\mathrm{SNIa}}(k',k) & \propto & \int\limits_{0}^{\tau_{\mathrm{G}}}w_{\mathrm{SNIa}}(k',t) w_{\mathrm{SNII}}(k-k',t)\times\nonumber\\ 
& \times & a_{\star}(t)w_{\mathrm{ISM}}(k,t)\psi(t)\mathrm{d}t,
\label{sniadistr}
\end{eqnarray}

\noindent
where \mbox{$a_{\star}(t)w_{\mathrm{ISM}}(k,t)\psi(t)$} may be regarded as a weighting function. Given $f_{\mathrm{SNIa}}$, the absolute number density of stars enriched by $k$ SNe, of which $k'$ are SNe of Type~Ia, is \mbox{$n_k\times f_{\mathrm{SNIa}}(k',k)$}. For example, out of $274.1$ stars per kpc$^3$ enriched by $100$ SNe (i.e., Model A$^{\prime}$), $42.5$ kpc$^{-3}$ are enriched by $6$ Type~Ia SNe (see Table \ref{fsnia_tab}). The majority of the stars enriched by $100$ SNe are found to be more metal-rich than \mbox{$[\mathrm{Fe}/\mathrm{H}]=-2.5$}, since the Fe yield of the Type~Ia SNe is on the order of $10$ times that of the average Fe yield of Type~II SNe (e.g., Nomoto et al. 1997\nocite{nomoto97ia}). Here, we take the yield to be \mbox{$p_{\mathrm{Fe}}=7.49\times 10^{-1}~\mathcal{M_{\odot}}$} for all Type~Ia SNe (Iwamoto et al. 1999\nocite{iwamoto99}, Model W7).

\begin{center}
\begin{table}[t]
\caption{Number of contributing thermonuclear SNe$^{\mathrm{a}}$}
  \label{fsnia_tab}
  \begin{tabular}{rcrcrc}
     \hline
     \hline
     \\*[-0.5em]
\footnotesize{$k'$} & \footnotesize{$f_{\mathrm{SNIa}}$} & \footnotesize{$k'$} & \footnotesize{$f_{\mathrm{SNIa}}$} & \footnotesize{$k'$} & \footnotesize{$f_{\mathrm{SNIa}}$} \\
     \\*[-0.5em]
     \hline
     \\*[-0.8em]
\footnotesize{$0$} & \footnotesize{$8.18\!\times\! 10^{-4}$} & \footnotesize{$7$} & \footnotesize{$1.54\!\times\! 10^{-1}$} & \footnotesize{$14$} & \footnotesize{$5.25\!\times\! 10^{-3}$}\\*[0.1em]
\footnotesize{$1$} & \footnotesize{$6.01\!\times\! 10^{-3}$} & \footnotesize{$8$} & \footnotesize{$1.32\!\times\! 10^{-1}$} & \footnotesize{$15$} & \footnotesize{$2.24\!\times\! 10^{-3}$}\\*[0.1em]
\footnotesize{$2$} & \footnotesize{$2.19\!\times\! 10^{-2}$} & \footnotesize{$9$} & \footnotesize{$1.00\!\times\! 10^{-1}$} & \footnotesize{$16$} & \footnotesize{$8.87\!\times\! 10^{-4}$}\\*[0.1em]
\footnotesize{$3$} & \footnotesize{$5.25\!\times\! 10^{-2}$} & \footnotesize{$10$} & \footnotesize{$6.75\!\times\! 10^{-2}$} & \footnotesize{$17$} & \footnotesize{$3.27\!\times\! 10^{-4}$}\\*[0.1em]
\footnotesize{$4$} & \footnotesize{$9.38\!\times\! 10^{-2}$} & \footnotesize{$11$} & \footnotesize{$4.09\!\times\! 10^{-2}$} & \footnotesize{$18$} & \footnotesize{$1.13\!\times\! 10^{-4}$}\\*[0.1em]
\footnotesize{$5$} & \footnotesize{$1.33\!\times\! 10^{-1}$} & \footnotesize{$12$} & \footnotesize{$2.25\!\times\! 10^{-2}$} & \footnotesize{$19$} & \footnotesize{$3.64\!\times\! 10^{-5}$}\\*[0.1em]
\footnotesize{$6$} & \footnotesize{$1.55\!\times\! 10^{-1}$} & \footnotesize{$13$} & \footnotesize{$1.13\!\times\! 10^{-2}$} & \footnotesize{$20$} & \footnotesize{$1.10\!\times\! 10^{-5}$} \\
     \\*[-0.8em]
     \hline
     \\*[-0.8em]
  \end{tabular}

\hspace*{5pt}\footnotesize{$^{\mathrm{a}}$Model A$^{\prime}$ for stars enriched by $k=100$ SNe} \\
\end{table}
\end{center}

\begin{figure}[t]
 \resizebox{\hsize}{!}{\includegraphics{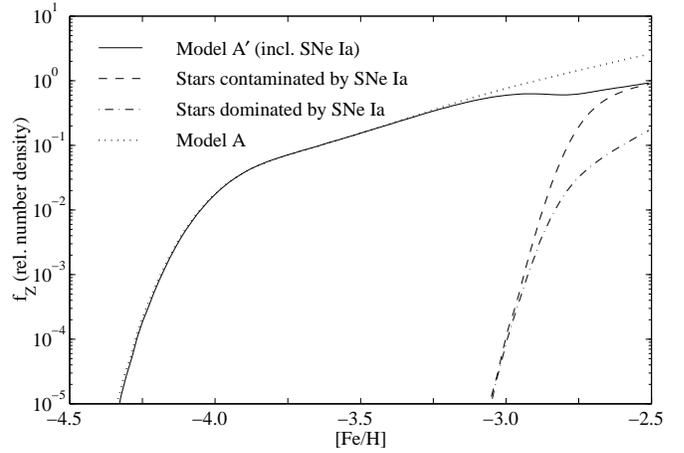}}
 \caption{The effect of thermonuclear (Type~Ia) SNe on the metallicity distribution of extremely metal-poor stars. The metal-poor tail is completely unaffected by Type~Ia SNe. The fraction of stars contaminated by one or more Type~Ia SNe (dashed line) only becomes significant above $[\mathrm{Fe}/\mathrm{H}]\sim -2.8$. The fraction of stars with $>50\%$ of their Fe produced in Type~Ia SNe (dash-dotted line) remains small, even at $[\mathrm{Fe}/\mathrm{H}]=-2.5$. The metal-rich end of Model A$^{\prime}$ (full line) is depleted in stars, as compared to Model A (dotted line), as these stars are pushed towards higher metallicities due to the high Fe yield of the Type~Ia SNe.}
 \label{modelAsnia_fig}
\end{figure}

\subsubsection{The mixing mass distribution for thermo- nuclear SNe}
\noindent
Since the onset of Type~Ia SNe is delayed relative to the Type~II SNe, the mixing masses of the Type~Ia SNe are, on average, smaller than the corresponding mixing masses of the Type~II SNe at a given time $t$ or for a given total number $k$ of enriching SNe. Since all mixing volumes are assumed to grow at the same speed, the mixing mass distribution for the Type~Ia SNe are, as for the Type~II SNe, governed by Eq. (\ref{generalfm}) or Eq. (\ref{derivedfm}), however, with the SN rate \mbox{$u_{\mathrm{SN}}(t-\tau_{V})$} replaced by the appropriate rate $u_{\mathrm{SNIa}}$. Similarly, the mixing mass distribution for the Type~II SNe is now determined by the rate $u_{\mathrm{SNII}}$.

\subsubsection{The effect of thermonuclear SNe}
\noindent
Fig. \ref{modelAsnia_fig} shows that the metallicity distribution is unaffected by the presence of Type~Ia SNe below \mbox{$[\mathrm{Fe}/\mathrm{H}]\simeq -3$} for this particular choice of input model and $u_{\mathrm{SNIa}}$. The metal-rich end of Model A$^{\prime}$ is, however, depleted in stars, as compared to Model A. These ``missing'' stars have been enriched by Type~Ia SNe to such an extent that they end up with metallicities higher than \mbox{$[\mathrm{Fe}/\mathrm{H}]= -2.5$}. Although almost all the remaining stars at \mbox{$[\mathrm{Fe}/\mathrm{H}]=-2.5$} may be contaminated by debris from Type~Ia SNe (about $95\%$ in Model A$^{\prime}$), the Fe in the large majority of these stars is produced predominantly in the core-collapse SNe (see the dash-dotted curve in Fig. \ref{modelAsnia_fig}). If the rate of Type~Ia SNe was significantly higher in this early phase and if the ISM was heavily affected by infall of pristine gas (e.g., as in Model C), thermonuclear SNe might have been the dominant source of Fe even at metallicities as low as \mbox{$[\mathrm{Fe}/\mathrm{H}]=-2.5$}. This is, however, at variance with the general belief that Type~Ia SNe did not become an important source of Fe until $[\mathrm{Fe}/\mathrm{H}]\sim-1$, as suggested by observations of stars in the Galactic disk. In order to achieve such a ``late'' onset of the Type~Ia SNe, a much higher SFR, a much lower density, or a much lower mixing velocity $\sigma_{\mathrm{mix}}$ than adopted would be required in the early Galaxy. Note, however, that the low-metallicity cut-off would then be displaced accordingly. Alternatively, no (or very few) Type~Ia SN explosions occurred in the early Galaxy. In fact, there are studies indicating that the formation rate of Type~Ia SN precursor systems is metallicity-dependent and that this rate may be negligible at metallicities below \mbox{$[\mathrm{Fe}/\mathrm{H}]=-1$} (Kobayashi et al. 1998\nocite{kobayashi98}). In conclusion, extremely metal-poor stars in the Galactic halo should, in general, be affected very little by Type~Ia SNe. 

\begin{figure}[t]
 \resizebox{\hsize}{!}{\includegraphics{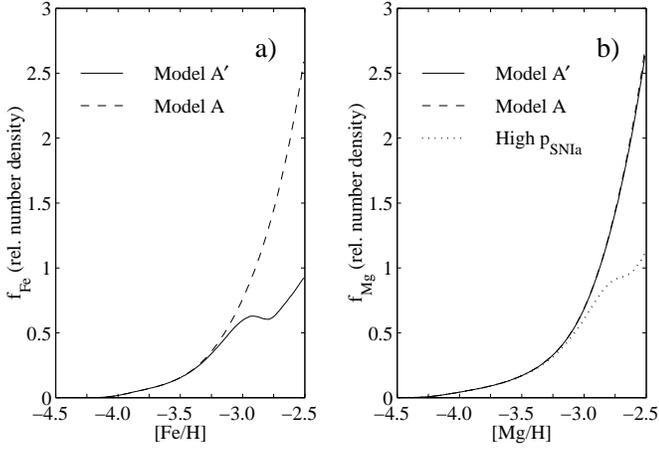}}
 \caption{The difference between high and low \mbox{$p_{\mathrm{SNIa}}/p_{\mathrm{SNII}}$} yield ratios. {\bf a)} Metallicity distributions for Fe with (Model A$^{\prime}$, full line) and without (Model A, dashed line) the presence of Type~Ia SNe. The onset of the Type~Ia SNe is clearly seen. {\bf b)} The corresponding metallicity distributions, however, with Mg as metallicity indicator. The onset of the Type~Ia SNe is virtually undetectable due to the low $p_{\mathrm{SNIa}}=8.57\times 10^{-3}~\mathcal{M_{\odot}}$ of Mg (Iwamoto et al. 1999, Model W7). The dotted line denotes the metallicity distribution for a $68$ times higher $p_{\mathrm{SNIa}}$ of Mg, yielding a $p_{\mathrm{SNIa}}/p_{\mathrm{SNII}}$ ratio similar to that of Fe.}
 \label{metdistr}
\end{figure}

\par

Let us take a closer look at the ``missing'' stars. As mentioned above, these stars are pushed towards higher metallicities due to the presence of the Type~Ia SNe. In fact, about $65\%$ of all stars are missing at \mbox{$[\mathrm{Fe}/\mathrm{H}]= -2.5$} in comparison to Model A. Apparently, when the Type~Ia SNe kick in, they seem to kick in hard. The effect is caused by the high SNIa/SNII yield ratio of Fe \mbox{($p_{\mathrm{SNIa}}/p_{\mathrm{SNII}}\sim 9$)}, which creates a hole (it actually looks more like a bump when plotted on a linear scale, see Fig. \ref{metdistr}a) in the metallicity distribution measured by Fe. In contrast to holes and bumps created by, e.g., variations in the infall rate of pristine gas, this feature will not have a distinct counterpart in distributions for elements with a lower SNIa/SNII yield ratio, such as Si \mbox{($p_{\mathrm{SNIa}}/p_{\mathrm{SNII}}\sim 2$)}, Sc \mbox{($p_{\mathrm{SNIa}}/p_{\mathrm{SNII}}\sim 1$)}, or even better Mg \mbox{($p_{\mathrm{SNIa}}/p_{\mathrm{SNII}}\lesssim 0.1$}, the yield ratios are estimated from data published by Iwamoto et al. 1999\nocite{iwamoto99}). This effect is illustrated in Fig. \ref{metdistr}. It is clear that the onset of the Type~Ia SNe is imprinted differently in metallicity distributions of different elements. An Fe hole/bump feature with no counterpart in the distribution for, say, Mg, thus suggests an additional contribution of Fe from Type~Ia SNe. Such a feature may therefore be used to determine whether Galactic halo stars with relatively low $[\alpha/\mathrm{Fe}]$ ratios have been enriched by Type~Ia SNe. Note, however, that for Mg in particular, the metal-rich end of the distribution may be affected by intermediate-mass stars.

\subsection{Comparison to Oey (2000)} \label{results_procon}
\noindent
Although developed independently, our mathematical description of the evolution of chemical inhomogeneities in the ISM resembles that of Oey (2000\nocite{oey00}); nevertheless, the two approaches describe two physically different pictures of the chemical enrichment in metal-poor systems. We view the evolution of chemical inhomogeneities in terms of mixing volumes generated by individual SNe. Oey, on the other hand, uses the concept of star-forming regions with a fixed distribution of sizes $V_s(N_{\ast})$ of the star-forming regions, where the number $N_{\ast}$ of SNe within each star-forming region is estimated from the mechanical luminosity function, assumed to be a power law with index $-\beta$. The metallicity of each star-forming region then depends on its predetermined size and is given by $N_{\ast}\times m_y/V_s(N_{\ast})$, where $m_y$ is the mean yield of metals per SN. Conversely, in our approach the size of each mixing volume (or mixing mass) grows with time, and the metallicity within each volume is determined by the mass-dependent yield from a single SN. This enables us to follow the evolution of chemical inhomogeneities caused by individual SNe. 

\par

Oey assumes that star-forming regions are uncorrelated and randomly distributed in space, meaning that the binomial distribution can be used to describe the probability of finding a point in the ISM occupied by $k$ (Oey uses $j$) overlapping regions. We assume a spatially random distribution as well, but in relation to mixing volumes (i.e., uncorrelated star formation). Although we use the Poisson distribution instead of the binomial distribution, the difference is negligible since the average number $\mu$ of enriching SNe is much smaller than the total number of SNe in the system. Mathematically, our parameter $\mu$ corresponds to Oey's parameter $nQ$, the number of generations of star-forming regions times one volume-filling factor per generation. However, the two are physically different in that Oey assumes that star formation and SN enrichment occurs in local regions. Our description can be modified in this direction by replacing our mixing volumes with star-forming regions (see Paper~III\nocite{paperiii}). However, we note that our model is not closed; i.e., we allow for the gas density to vary with time.

\subsection{Inhomogeneous infall: The next step} \label{results_infall}
\noindent
By treating the infall as a diffuse flow we tend to underestimate the amplitude of the chemical inhomogeneities. The initial and maximum decrease in the local abundance due to an infalling cloud of pristine gas is determined by the density contrast $\rho_{\mathrm{c}}/\rho$, where $\rho_{\mathrm{c}}$ is the density of the cloud. This ratio does not depend on the metallicity of the ISM. On the other hand, the abundance increase due to an exploding SN is measured by the ratio between the yield $p$ of the element and the mass of the element already present in the ISM within the volume affected by the SN. This ratio can be written as $p/\epsilon M_{\mathrm{mix}}^{\mathrm{SN}}$, where $\epsilon$ is the mass fraction of the element and $M_{\mathrm{mix}}^{\mathrm{SN}}$ is the total mass within the affected volume. Comparing these two ratios we see that for mixing masses $M_{\mathrm{mix}}^{\mathrm{SN}} < (p/\epsilon)/(\rho_{\mathrm{c}}/\rho)$, the inhomogeneities in the ISM are dominated by those from the SNe. Using this relation with, e.g., $p_{\mathrm{Fe}}=0.1~\mathcal{M_{\odot}}$, $\epsilon_{\mathrm{Fe}}=10^{-6}$ (corresponding to an abundance $[\mathrm{Fe}/\mathrm{H}]=-3$), and $\rho_{\mathrm{c}}/\rho=1$, we obtain $M_{\mathrm{mix}}^{\mathrm{SN}} < 10^5~\mathcal{M_{\odot}}$. Thus, the large amplitude inhomogeneities in $[\mathrm{Fe}/\mathrm{H}]$ due to SNe are relatively small in size and, as shown in Fig.~\ref{Mmix_fig}, the SN material is typically mixed with a mass which is an order of magnitude larger before being locked up in subsequently formed stars. Conversely, if we assume that an infalling cloud of mass $M_{\mathrm{c}}$ is mixed within a volume of mass $M_{\mathrm{mix}}^{\mathrm{c}}$, the criterion for SN-dominated inhomogeneities is $M_{\mathrm{c}} < (p/\epsilon)/(M_{\mathrm{mix}}^{\mathrm{SN}}/M_{\mathrm{mix}}^{\mathrm{c}})$. Taking the ratio of the mixing masses to be on the order of unity, $M_{\mathrm{c}} < 10^{5}~\mathcal{M_{\odot}}$ for the values of $p_{\mathrm{Fe}}$ and $\epsilon_{\mathrm{Fe}}$ given above. Again, this is a relatively small mass, in this case for the infalling clouds. Although the chemical inhomogeneities are dispersed in the mixing process as well, they are probably affected by heterogeneous infall, increasing the star-to-star scatter in abundances relative to hydrogen and displacing the cut-off towards lower metallicities. Since a spatially varying density $\rho(t,\mathbf{x})$ would have a similar effect on the development of chemical inhomogeneities as infalling gas clouds, the treatment presented in this study gives a lower limit to the chemical abundance scatter in stars. However, relative abundance ratios of two heavier elements are not in themselves affected by how much the gas is diluted. Diagrams relating abundance ratios of heavier elements, so-called $A/A$ diagrams, are therefore less sensitive to the mixing uncertainties than $A/\mathrm{H}$ diagrams, in which the abundance of one element is measured relative to hydrogen. 

\par

We note that the present theory may be naturally extended to allow for the stochastic infall of gas clouds, in an approach similar to that used here for the description of chemical inhomogeneities caused by SNe.    

\section{Summary}\label{conclusions}
\noindent
The Poisson distribution $e^{-\mu(t)}\mu(t)^k/k!$ is taken to describe the development chemical inhomogeneities in the ISM caused by randomly distributed SNe. If the time-like parameter $\mu(t)$ denotes the average number of SNe that have enriched a random volume element in space at time $t$, the Poisson distribution gives the probability of finding a point in space enriched by $k$ SNe at time $t$. Defining the mixing volume $V_{\mathrm{mix}}(t)$ to be the volume at time $t$ affected by material ejected in a single SN explosion allows us to account for large-scale mixing by, e.g., turbulent motions. This mixing volume is then used in the calculation of $\mu(t)$. Infall is treated as a homogeneous, diffuse flow onto the system, affecting the overall gas density $\rho(t)$. 

\par

An analytical expression for the number density $n_k$ of stars enriched by a given number $k$ of SNe is derived. It is shown that the ratio $\psi/\frac{\mathrm{d}\mu}{\mathrm{d}t}$, where $\psi$ is the SFR and $\frac{\mathrm{d}\mu}{\mathrm{d}t}$ is the rate of change of $\mu$, is a measure of $n_k$. An expression for the probability density function $f_{M_k}$ of mixing masses is also given. This function describes the probability of how much the ejected material from a SN is diluted (i.e., mixed with a certain mixing mass $M$), before being locked up in a subsequently formed star. This distribution depends on the number $k$ of SNe that have enriched the star.

\par

These results are combined to calculate the metallicity distribution of low-mass stars in the Galactic halo. Three models with different values for the parameters $\psi(t)$ and $\rho(t)$  are discussed (see Sect.~\ref{appl_models}). They all show a steep decline in the number of stars below $[\mathrm{Fe}/\mathrm{H}]\simeq -4$, in agreement with observations. This cut-off is a natural consequence of the lower  probability of finding a large mixing volume enriched by a single SN, producing a small Fe yield. The cut-off is thus determined by the tail of $f_{M_{k=1}}$ (together with the yield). Fig.~\ref{nZ_fig} and Fig.~\ref{kdistr_fig} indicate that above the cut-off there is a high probability of finding a star enriched by several SNe, as concluded by Nissen et al.\ (1994)\nocite{netal94} and more recently also suggested, e.g., by Cayrel et al.\ (2004\nocite{cayrel04}). In fact, only the stars below the cut-off have a high probability ($\sim 1$) of being enriched by individual SNe. Note, however, that about one percent of all the stars enriched by one single SN have a metallicity $[\mathrm{Fe}/\mathrm{H}]>-2.5$ (see Table~\ref{restable}). 

\par

The absolute number density $n_{\mathrm{III}}$ of metal-free Population~III stars is directly given by Eq.~(\ref{Nk}), or alternatively Eq.~(\ref{Nkexp}), for $k=0$. The fraction of metal-free stars is estimated to be $(1-2)\times 10^{-2}$, which is approximately two orders of magnitude higher than the current observational value of $4\times 10^{-4}$ (cf. Oey 2003\nocite{oey03}). This discrepancy was also found by Oey (2003\nocite{oey03}). However, if the formation of low-mass Population III stars was inhibited due to inefficient cooling of the primordial ISM, interrupting further fragmentation (e.g., Bromm et al.\ 1999)\nocite{bromm99}, the fraction of stars below $[\mathrm{Fe}/\mathrm{H}]= -4$ is brought down to $10^{-4}-10^{-3}$, in agreement with the current observational estimate. 

\par

Even in the short formation time-scale scenario, and for a relatively high rate $u_{\mathrm{SNIa}}$, the contaminating effect of thermonuclear SNe is found to be insignificant over the entire extremely metal-poor regime with a possible exception at the very metal-rich end. However, although the majority of stars may be contaminated by Type~Ia SNe at $[\mathrm{Fe}/\mathrm{H}]\sim -2.5$, the fraction of stars with more than $50\%$ of their Fe produced in the Type~Ia SNe remains small. Stars that are heavily polluted by Type~Ia SNe are pushed towards higher metallicities, thereby creating a hole/bump in the metallicity distribution. This effect is pronounced due to the high yield ratio $p_{\mathrm{SNIa}}/p_{\mathrm{SNII}}$ of Fe. Such a feature would be virtually absent in distributions for elements with much lower $p_{\mathrm{SNIa}}/p_{\mathrm{SNII}}$, e.g., for Mg. Hence, the existence of an Fe hole/bump feature with no counterpart in, e.g., Mg would indicate the presence of a subpopulation of Galactic halo stars that have been enriched by Type~Ia SNe. 

\par

Finally, we note that an inhomogeneous infall in the form of gas clouds, pristine as well as pre-enriched, would increase the amplitude of the chemical inhomogeneities and most probably increase the star-to-star scatter in the abundances relative to hydrogen. The effect of infalling clouds can be treated stochastically and should be possible to incorporate in an extended version of the present theory.

\begin{acknowledgements}
\noindent
I'm debted to Bengt Gustafsson for numerous enlightening discussions on stochastic chemical enrichment and for scrutinizing the manuscript in detail. Valuable comments on the manuscript were also made by Michelle Mizuno-Wiedner, Lee Ann Willson, and the anonymous referee. This work was supported by the Swedish Research Council.  
\end{acknowledgements}

\bibliographystyle{bibtex/apj}
\bibliography{ref}

\end{document}